\documentclass{IEEEtran}
\usepackage{float}
\usepackage{amsmath}
\usepackage{amsthm}
\usepackage{amssymb}	
\usepackage{graphicx}
\usepackage{subfigure}
\usepackage{enumitem}
\usepackage{algorithm}
\newcommand*{\J}{\jmath}%
\usepackage{amsmath,amssymb,mathtools,bm,etoolbox}
\newtheorem{my_theorem}{Theorem}

\newtheorem{my_lemma}{Lemma}

\newtheorem{my_proposition}{Proposition}

\usepackage{color}
\usepackage{mathtools}
\usepackage{suffix}
\usepackage{breqn}
\usepackage{hhline}
\usepackage{cite}

\DeclarePairedDelimiterXPP\Aver[1]{\mathbb{E}}{[}{]}{}{
	
	#1
}
\DeclarePairedDelimiterX\MeijerM[3]{\lparen}{\rparen}%
{\,#3\delimsize\vert\begin{matrix}#1 \\ #2\end{matrix}}
\newcommand\MeijerG[8][]{%
	G^{\,#2,#3}_{#4,#5}\MeijerM[#1]{#6}{#7}{#8}}
\newcommand{\diff}{\mathop{}\!d}

\WithSuffix\newcommand\MeijerG*[7]{%
	G^{\,#1,#2}_{#3,#4}\MeijerM*{#5}{#6}{#7}}

\title{Unified Performance Analysis of  Reconfigurable Intelligent Surface Empowered  Free-Space Optical Communications} 
\author{Vinay Kumar Chapala,~\IEEEmembership{Graduate Student Member,~IEEE} and  S.~M.~Zafaruddin,~\IEEEmembership{Senior Member,~IEEE} 
	
	\thanks{Vinay Kumar Chapala (p20200110@pilani.bits-pilani.ac.in) and S.~M.~Zafaruddin (syed.zafaruddin@pilani.bits-pilani.ac.in)  are  with  the Department of Electrical and Electronics Engineering, Birla Institute of Technology and Science, Pilani, Pilani-333031, Rajasthan, India.}
	\thanks{ This work was supported in part by the Science and
		Engineering Research Board (SERB), Department of Science and Technology
		(DST), Government of India, under Start-up Research Grant SRG/2019/002345.}
}

\begin{document}
	\maketitle
	\begin{abstract}
		Reconfigurable intelligent surface (RIS) is an excellent use case for line-of-sight (LOS) based technologies such as free-space optical (FSO) communications.  In this paper, we analyze the performance of RIS-empowered FSO (RISE-FSO) systems by unifying Fisher–Snedecor (${\cal{F}}$), Gamma-Gamma ($\cal{GG}$), and Mal\'aga ($\cal{M}$) distributions for atmospheric turbulence with zero-boresight pointing errors over deterministic as well as  random path-loss in foggy conditions with heterodyne detection (HD) and intensity modulation/direct detection (IM/DD) methods. By deriving the probability density function (PDF) and cumulative distribution function (CDF) of the direct-link (DL) with the statistical effect of atmospheric turbulence, pointing errors and random fog, we develop exact expressions of PDF and CDF of the resultant channel for the RISE-FSO system. Using the derived statistical results, we present exact expressions of outage probability, average bit-error-rate (BER), ergodic capacity, and moments of signal-to-noise  ratio (SNR) for both  DL-FSO and		 RISE-FSO systems. We also develop an asymptotic analysis of the outage probability and average BER and derive the diversity order of the considered systems. We validate the analytical expressions using Monte-Carlo simulations and demonstrate the performance scaling of the FSO system with the number of  RIS elements for various turbulence channels, detection techniques, and weather conditions.
		
	\end{abstract}
	
	\section{Introduction}
	Reconfigurable intelligent surface (RIS) is a promising technology to empower wireless systems by artificially controlling the characteristics of propagating signals  in a desired direction \cite{Renzo2019,Basar2019_access,ElMossallamy2020, Qingqing2021,Yuan21}. Specifically,  RISs are constructed by planar metasurfaces using  a large number of reflection units  adapted by integrated electronics to control the phase, amplitude and polarization of  incident signals. The RIS is a promising  alternative to active relying techniques  without requiring complex processing at the relay to improve the performance of wireless systems. Free-space optical (FSO) communication is a potential technology to cater high data rate  transmission  with license-free operation over a huge bandwidth in the optical spectrum \cite{Khalighi2014}. Comparing with radio-frequency (RF), FSO systems are immune to the electromagnetic  interference and have been considered  a  cost-effective solution  for  terrestrial backhaul/fronthaul wireless applications for 5G and beyond 5G networks \cite{Yang2019,Dang2020}.
	
	The FSO link is subjected to various channel impairments such as atmospheric turbulence, pointing errors, and other weather conditions. The atmospheric turbulence is  the scintillation effect of light propagation and introduces fading in the transmitted signal. In addition to the atmospheric turbulence, the range of FSO link is limited   due to the higher signal  attenuation, especially in the presence of fog and dust.  Moreover, FSO is a line-of-sight (LOS) technology that may suffer significant performance degradation in the presence of pointing errors caused by the misalignment between  the transmitter and the receiver. In this context, the deployment of FSO is not feasible for terrestrial applications due to the unavailability of a direct link in the presence of obstructions creating dead zones for wireless connectivity. The use of cooperative  relaying  has  been extensively studied to mitigate the effect of channel impairments to improve the performance of FSO  systems \cite{Safari2008, Ansari2016, dual_hop_turb2017, Zhang2020}.

	The advent of RIS opens an exciting research avenue to investigate and improve the performance of wireless systems for ubiquitous connectivity.  Recently, the performance of RIS-enabled wireless systems have been analyzed over radio-frequency (RF) transmissions \cite{Kudathanthirige2020, Boulogeorgos2020_ris, Boulogeorgos2020_access,  Liang2020, Qin2020, Ferreira2020,   Selimis2021, Ibrahim2021_tvt,trigui2020_fox, du2021}, mixed RF-FSO  \cite{Liang2020_vlc,Yang_2020_fso,Sikri21},  and FSO  systems \cite{Jamali2021, Najafi2019, Wang2020, yang2020fso,ndjiongue2021}.  In \cite{Liang2020_vlc}, a RIS-assisted dual-hop visible light communication (VLC)-RF system for an indoor scenario was proposed with VLC in the first link and  RIS in the second RF link. In \cite{Yang_2020_fso}, the authors considered a decode-and-forward (DF) relaying  to mix an FSO link over Gamma-Gamma turbulence with pointing errors  and RF link assisted with RIS over Rayleigh fading.  In a similar setup, the mixed RF-FSO system was analyzed in  \cite{Sikri21} by considering an additional co-channel interference (CCI) in the RF link. The use of relaying in such systems decouples the performance analysis for FSO and RF such that the impact of RIS is present  only in the RF link without the challenges of analyzing the RIS  for  FSO transmissions with complicated fading models. Recently, the authors in \cite{Jamali2021, Najafi2019,Wang2020, yang2020fso,ndjiongue2021}  employ RIS  module for FSO systems.  An overview of  various design aspects of optical RIS for FSO comparing with RIS-assisted RF is presented  \cite{Jamali2021}. The authors in \cite{Najafi2019} characterized the impact of the physical parameters of the RIS to model the geometric and misalignment losses due to the random movements of the RIS and the effect of building sway. In \cite{Wang2020}, multiple optical RISs are used to improve the outage probability of the FSO system under the effect of  pointing errors  without considering the atmospheric turbulence. The authors in  \cite{yang2020fso,ndjiongue2021} considered the  Gamma-Gamma atmospheric turbulence with pointing errors to analyze the RIS based FSO system. However, the authors in \cite{ndjiongue2021}  considered a simplified model by considering a single-element RIS to assist the FSO system. Moreover, the authors in \cite{yang2020fso} used the Gaussian distribution to analyze the RIS-assisted FSO system by employing the central limit theorem to approximate the distribution function of the end-to-end channel. It is desirable to provide an exact analysis of the  FSO system with atmospheric turbulence  combined with pointing errors and assisted by  the RIS with multiple elements. To this end, we emphasize that the RIS can be applied  for both indoor and outdoor FSO communications. In indoor  applications, when there is no direct link between the source and destination, the use of RIS (deployed on the ceiling or wall) can reflect the incident laser beam to the receiver coherently. In the devoid of RIS, the intensity of  diffused light scattered from rough surfaces can be much lower for establishing the communication link \cite{Cao2019}.  The optical RIS module can also be mounted on the top of a building   for outdoor applications, facilitating building to building (B2B) connection for high-speed backhaul links.  
	
	There are several statistical models in the literature to characterize the atmospheric turbulence depending on the severity of turbulence, type of wave propagation, and mathematical tractability of the model. The Gamma-Gamma ($\cal{GG}$) is widely accepted for moderate-to-strong turbulence regime \cite{Habash2001}, whereas the  generalized Mal\'aga model ($\cal{M}$) can be used for all irradiance conditions in homogeneous and isotropic turbulence \cite{malaga2011}.  Recently, the Fisher–Snedecor $\cal{F}$-distribution model for the atmospheric turbulence is proposed for its mathematical tractability \cite{Peppas2020}. On the other hand, the zero-boresight  model proposed by \cite{Farid2007}  is widely used in the literature to characterize the pointing errors in FSO systems.   Traditionally, signal attenuation for FSO transmissions is assumed to be deterministic and quantified using a visibility range, for example, less attenuation in haze and light fog and more loss of signal power in the dense fog \cite{Kim2001}. However, recent measurement data confirm that the signal attenuation  in the fog is not deterministic but follows a probabilistic model  \cite{Esmail2017_Access}. It should be noted that heterodyne detection (HD) and intensity modulation/direct detection (IM/DD) are the two main modes of detection in FSO systems.   Considering such a diverse operation,  it is desirable to unify the FSO  using different models of atmospheric turbulence with pointing errors, path loss, and  detection modes.

	In this paper,  we analyze the performance of a RIS-empowered FSO (RISE-FSO) system by unifying ${\cal{F}}$, ${\cal{GG}}$, and $\cal{M}$ distributions for atmospheric turbulence with zero-boresight pointing errors over deterministic as well as  random path-loss in foggy conditions with HD and IM/DD modes of detection. It is emphasized that such a unification is not straight forward and it is not available even for the direct-link (DL)  FSO systems.
	The major  contributions of the proposed work are summarized as follows:
	\begin{itemize}[leftmargin=*]
		\item We derive the probability density function (PDF) and  cumulative distribution function (CDF) of the combined statistical effect of random fog with atmospheric turbulence and  pointing errors of a DL-FSO system   by unifying ${\cal{F}}$, ${\cal{GG}}$, and $\cal{M}$ atmospheric turbulence models such that the traditional deterministic path loss model remains a particular case for a unified performance analysis.
		
		\item To analyze the RISE-FSO, we derive exact closed form expressions of PDF and CDF for the resultant channel realized by  the sum of products (SOP) of  fading coefficients considered to be  independent but not identically distributed (i.ni.d) according to the DL-FSO fading channel.   
		
		\item Using the derived PDF and CDF, we analyze the performance of RISE-FSO system by developing exact closed-form expressions of the  outage probability, average bit-error-rate (BER), ergodic capacity,  and moments of signal-to-noise ratio (SNR) in terms of Fox's H-function.   For  comparison, we also  develop an exact analysis of the aforementioned performance metrics for the DL-FSO system, which is not available under the combined effect of atmospheric turbulence, pointing errors, and  random fog.
		
		\item We present  asymptotic analysis for the outage probability and average BER in  simpler  Gamma function in the high SNR regime. The asymptotic expressions  are readily tractable and provide engineering insights for system design. As such, we derive diversity order using the outage probability and average BER depicting   the impact of  atmospheric turbulence, pointing errors, and foggy channel on system behavior,  and the  scaling of FSO performance  with an increase in the number of RIS elements.

		\item We validate the derived analytical results using extensive Monte-Carlo simulations demonstrating the effectiveness of RISE-FSO in comparison with the DL-FSO system for various atmospheric turbulence, detection techniques, and weather conditions.
		
	\end{itemize}

	\subsection{Related Works}
In this subsection, we summarize recent research works on RIS based RF systems. The authors in  \cite{Kudathanthirige2020} analyzed outage probability, average bit-error-rate (BER) and
bounds on capacity over Rayleigh fading RIS system. Considering the similar channel model, the authors in \cite{Boulogeorgos2020_ris} presented the exact and asymptotic analysis of ergodic capacity. In \cite{Boulogeorgos2020_access}, performance of RIS-assisted and amplify-and-forward (AF) relaying wireless systems were compared for Rayleigh fading channel where the  RIS-assisted system was shown to outperform the corresponding relaying systems. In \cite{Liang2020}, for arbitrarily finite RIS elements, the authors offered closed-form estimates on the channel distribution over Rayleigh fading channels for dual-hop and transmit RIS-aided schemes. The authors used Rician fading to investigate the outage probability and ergodic capacity of a single-input single-output (SISO) RIS-assisted wireless communications system in \cite{Qin2020}. The authors in \cite{Ferreira2020,  Selimis2021}   approximated the average BER and ergodic capacity performance over Nakagami-m fading channels.  Exact coverage analysis of RIS-enabled systems with Nakagami-m channels was presented in \cite{Ibrahim2021_tvt}.  In \cite{trigui2020_fox}, the authors derived  exact expressions of the outage probability and ergodic capacity for a RIS-assisted system without pointing errors over generalized Fox’s H fading channels.  In \cite{Kong2021},  the authors considered $\alpha$-$\mu$ fading model and  analyzed the effective rate of RIS-assisted communications by simplifying the analysis using the  mixture of Gaussian instead of considering  the sum of cascaded $\alpha$-$\mu$ distributed random variables.
In \cite{du2021}, the authors analyzed a  RIS-assisted millimeter-wave communication over the fluctuating two rays (FTR) fading model in terms of Fox's H-function. Similarly,  the authors in \cite{du2020_thz} extended the analysis  for RIS-aided THz communications by deriving outage probability and ergodic capacity over FTR channel model combined with antenna misalignment  and hardware impairments. Recently, the performance of RIS-assisted THz transmissions over $\alpha$-$\mu$ fading channel with pointing errors is analyzed \cite{chapala2021THz}. The research on RIS-assisted wireless systems is growing rapidly. To the best of authors' knowledge,  exact  performance analysis on RIS empowered FSO system over generalized atmospheric turbulence with pointing errors is not publicly available.

	\subsection{Notations and Organizations}
Main notations used in this paper are summarized in Table \ref{table:notation_parameters}. The paper is organized as follows: system and channel models are summarized in Section II followed by statistical distribution functions of DL-FSO and RISE-FSO systems in Section III. Performance analysis through exact and asymptotic expressions is presented in Section IV. The numerical and simulation results are discussed in Section V. Finally, the paper concludes with Section VI.

\begin{table*}[tp]
	
		\caption{List of Main  Notations  }
		\label{table:notation_parameters}
		\centering
		\begin{tabular}{ c c c c }
			\hline
			\hline
			Notation   & Description & Notation & Description \\
			\hline
			$(\cdot)^{(f)}$ &  Fog & $\J$ & Imaginary number \\  \hline
			$(\cdot)^{(t)}$ & Atmospheric turbulence & $\Aver{.}$ &Expectation operator \\ \hline
			$(\cdot)^{(p)}$ &  Pointing errors & $\exp(.)$ &  Exponential function \\ \hline
			$(\cdot)^{(\rm tp)}$ &  Turbulence with pointing error & $(\cdot)^{(\rm DL)}$ & Direct link from source to destination \\ \hline
			$G_{p,q}^{m,n}  \left(x \middle\vert \begin{array}{c} a\\ b \end{array} \right)$ & Meijer's G-function & $H_{p,q}^{m,n}  \left(x \middle\vert \begin{array}{c} (a,A)\\ (b,B) \end{array}\right)$ & Single-variate Fox's H-function \\ \hline
			$\{a_{i}\}_{1}^{N} = \{a_{1},\cdots,a_{N}\}$ & Shorthand notation & $\Gamma(a)$ & $\int_{0}^{\infty}u^{a-1} e^{-u} \diff u$ \\
			\hline
			\hline
			
		\end{tabular}
\end{table*}

\section{System Model}	
\begin{figure}[tp]
	\centering	{\includegraphics[scale=0.5]{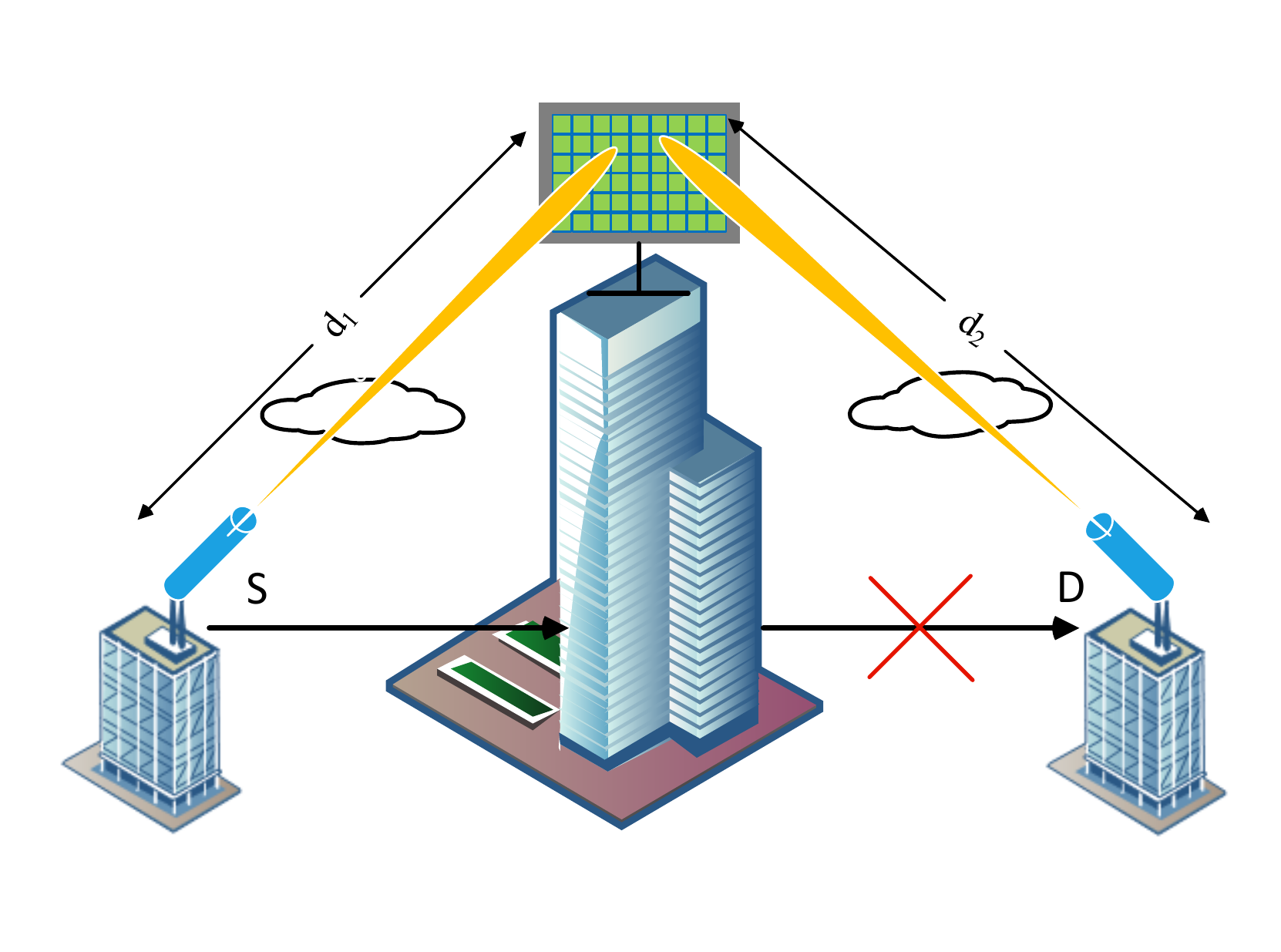}}
	\vspace{-2mm}
	\caption{Schematic diagram of RISE-FSO system for a typical outdoor application.}
	\label{fig:system_model}
\end{figure}

We consider a single-aperture FSO system where 	the  source $S$ wishes to communicate with the  destination $D$. We  assume that there is no direct link between the source and destination. To facilitate  transmissions for the RISE-FSO, we employ an $N$-element optical RIS such that a LOS exists from source to the  RIS and RIS to the destination, as shown in Fig. \ref{fig:system_model}.
Assuming perfect phase compensation at the RIS, the signal received  at the destination through RIS is expressed as \cite{yang2020fso}
\begin{equation}\label{model_smpl}
	y = \sum_{i=1}^{N} h_i g_i s + \nu
\end{equation}
where  $s$ is the transmitted signal with power $P_{T}$,   $h_i$  and  $g_i$  are channel fading coefficients between the source to the $i$-th RIS element and between the $i$-th RIS element to the destination, respectively, and $\nu$ is the additive Gaussian noise with variance $\sigma_\nu^2$. We denote by $d$ the link distance between the source and destination, by $d_1$ the distance between the source and the RIS,  and by $d_2$ the  distance between the RIS and destination.

We consider that FSO links experience signal fading due to  atmospheric turbulence, pointing errors, and  foggy conditions such that the combined fading  coefficient is denoted  as $h_i=h_i^{(f)} h_i^{(t)} h_i^{(p)}$  and $g_i=g_i^{(f)} g_i^{(t)} g_i^{(p)}$, where superscripts $^{(f)}$, $^{(t)}$, and $^{(p)}$ denote the fog, atmospheric turbulence, and pointing errors, respectively.  In what follows, we detail the modeling of the channel coefficient $h_i$. Note that we can model the channel coefficient $g_i$ similar to $h_i$. However,  we consider a general scenario considering  fading coefficients $h_i$ and $g_i$ to be independent  but  non-identical distributed (i.ni.d). Note that several publications employ  the assumption of  independent channels as a first approximation to analyze  RIS-assisted systems \cite{Liang2020, Boulogeorgos2020_access, Chongwen2019, Peng2021}.

To characterize the statistics of  pointing errors $h_i^{(p)}$, we use the recently proposed model for optical RIS in \cite{Wang2020}, which is based on the zero-boresight model \cite{Farid2007}:
\begin{equation}
	\begin{aligned}
		f_{h_i^{(p)}}(x) &= \frac{\rho^2}{A_{0}^{\rho^2}}x^{\rho^{2}-1},0 \leq x \leq A_0,
	\end{aligned}
	\label{eq:pdf_hp}
\end{equation}
where the term $A_0=\mbox{erf}(\upsilon)^2$ denotes the fraction of collected power. Define  $\upsilon=\sqrt{\pi/2}\ a_r/\omega_z$ with  $a_r$ as the aperture radius and  $\omega_z$ as the beam width.  We define the term $\rho^2 = {\frac{\omega^2_{z_{\rm eq}}}{\xi}}$ where $\omega_{z_{\rm eq}}$ is the equivalent beam-width at the receiver. The use of $\xi= 4 \sigma^2_{s}$ models the DL-FSO, where $\sigma_s^2$ is the variance of pointing errors displacement characterized by the horizontal sway and elevation \cite{Farid2007}, while $\xi= 4\sigma_{\theta}^2 d^2+16\sigma_{\beta}^2 d_2^2$ models the pointing errors for the  RIS-FSO system, where $\sigma_{\theta}$ and $\sigma_{\beta}$ represent pointing error and RIS jitter angle standard deviation defined in \cite{Wang2020}. It should be noted that the generalized non-zero boresight model \cite{Yang2014PE, Boluda-Ruiz:16} can also be considered to model pointing errors for RIS-assisted FSO systems. Although it would be interesting to analyze the  considered system with generalized pointing errors, we employ the zero-boresight model (as adopted in many reference papers) to avoid complicated analytical expressions. Further, the effect of non-zero  boresight and unequal jitter  is not significant on the performance of FSO systems \cite{Jung2020GPE}.

For a unified performance analysis over a variety of turbulence conditions,  we use the popular $\cal{GG}$ \cite{Habash2001}, the generalized $\cal{M}$ \cite{malaga2011},  and   recently introduced $\cal{F}$-distribution \cite{Peppas2020} to model the atmospheric turbulence.  We consider the   $\cal{M}$-distribution  since it is a generalized model  applicable for a wide range of atmospheric turbulence from weak to super strong in the saturation regime for plane and spherical waves propagation.  Further,  $\cal{GG}$ is the widely studied  model applicable for moderate to strong turbulence conditions.  Recently, the  $\cal{F}$-distribution is proposed as a computationally efficient  alternative to model various atmospheric turbulence for spherical and Gaussian propagation scenarios. We denote by $h_i^{(tp)}= h_i^{(t)}h_i^{(p)}$ the combined effect of atmospheric turbulence and pointing errors. Note that the product distribution of atmospheric turbulence and pointing errors is available in the literature \cite{Sandalidis2008_cl, Ansari2016, Peppas2020}. Thus, we represent the PDF of FSO link with   $\cal{GG}$ fading and  pointing errors as given in \cite{Sandalidis2008_cl}:
\begin{align}
		&f^{\rm GP}_{h_i^{(tp)}}(x)=\frac{\alpha_{{\scriptscriptstyle  G}} \beta_{{\scriptscriptstyle  G}}\rho^{2}}{A_0\Gamma(\alpha_{{\scriptscriptstyle  G}})\Gamma(\beta_{{\scriptscriptstyle  G}})}\nonumber\\&G_{1,3}^{3,0}\left(\frac{\alpha_{{\scriptscriptstyle  G}} \beta_{{\scriptscriptstyle  G}} x}{A_0} \left|
		\begin{array}{c}
			\rho^{2} \\
			\rho^{2}-1,\alpha_{{\scriptscriptstyle  G}}-1,\beta_{{\scriptscriptstyle  G}}-1
		\end{array}
		\right.\right)
		\label{eq:PDF_gg_pe}
\end{align}

where the fading parameters $\alpha_{{\scriptscriptstyle  G}}$ and  $\beta_{{\scriptscriptstyle  G}}$ are defined in \cite{Habash2001}. Similarly, the PDF of  $\cal{M}$-distributed turbulence combined with pointing errors is given as \cite{Ansari2016}:
\begin{align}
	\begin{split}
		f^{\rm MP}_{h_i^{(tp)}}(x)= \frac{\rho^{2}A_{\rm mg}}{2 x}\sum_{m=1}^{\beta_{{\scriptscriptstyle  M}}}b_{m}G_{1,3}^{3,0}\left(\frac{\alpha_{{\scriptscriptstyle  M}}\beta_{{\scriptscriptstyle  M}}}{g\beta_{{\scriptscriptstyle  M}}+\Omega^{'}}\frac{x}{A_{0}} \left|
		\begin{array}{c}
			\rho^{2}+1 \\
			\rho^{2},\alpha_{{\scriptscriptstyle  M}},m
		\end{array}
		\right.\right),
		\label{eq:PDF_Malaga_pe}
	\end{split}
\end{align}	
where the parameters  $\alpha_{{\scriptscriptstyle  M}}$, $\beta_{{\scriptscriptstyle  M}}$,  $A_{m\rm g}$, $b_m$, and $\Omega'$ are given in \cite{malaga2011}. Finally, the PDF of FSO channel experiencing ${\cal{F}}$-turbulence in the presence of pointing error impairments is given as \cite{Peppas2020}:
\begin{equation}
	f^{\rm FP}_{h^{(tp)}_i}(x)=\frac{\alpha_{{\scriptscriptstyle  F}}\rho^2G_{2,2}^{2,1}\left[\frac{\alpha_{{\scriptscriptstyle F}}}{(\beta_{{\scriptscriptstyle  F}}-1)A_0}x\left|\begin{array}{c}
			-\beta_{{\scriptscriptstyle  F}},\rho^2 \\
			\beta_{{\scriptscriptstyle  F}}-1,\rho^2-1\\
		\end{array}\right.\right]}{(\beta_{{\scriptscriptstyle  F}}-1)A_0\Gamma(\alpha_{{\scriptscriptstyle  F}})\Gamma(\beta_{{\scriptscriptstyle  F}})}
	\label{eq:PDF_F_pe}
\end{equation}
where  the fading parameters $\alpha_{{\scriptscriptstyle  F}}$ and  $\beta_{{\scriptscriptstyle  F}}$ are listed in \cite{Badarneh2020}.

The channel coefficient  $h_i^{(f)}$ models the path gain of signal transmission over the FSO link. Generally,  $h_i^{(f)}$ is a deterministic quantity obtained from Beer-Lambert's Law $h_i^{(f)}=e^{-\tau d_{}}$  where $d_{}$ is the link distance (in \mbox{km}) and $\tau$ is the atmospheric attenuation factor which depends  on the wavelength and visibility range \cite{Kim2001}. The atmospheric attenuation is defined as  $	\tau=\frac{3.19}{V}\left(\frac{\lambda}{550~\rm{nm}}\right)^{-q_v}$ where $V$ is the visibility (in \rm{km}), $\lambda$ is operating wavelength (in \rm{nm}), and $q_v$ is the size distribution of the scattering particles as  presented in \cite{Kim2001}. However, recent studies \cite{Esmail2017_Access, rahman2020_vinay} show that  the path gain in  foggy conditions exhibit randomness modeled with the following PDF
\begin{equation}
	f_{h_i^{(f)}}(x)= \frac{v^k}{\Gamma(k)}\left[\ln\left(\frac{1}{x}\right)\right]^{k-1} x^{v-1},
	\label{eq:h_fg}
\end{equation}
where $0<x\leq 1$, $v=4.343/d \beta^{{\scriptscriptstyle \rm fog}} $, $k>0$ is the shape parameter,  and $\beta^{{\scriptscriptstyle \rm fog}}>0$ is the scale parameter.

We also consider the DL-FSO system by considering the existence of a direct link between the source and destination as a means to compare with the RISE-FSO system: 
\begin{equation}\label{model_dl}
	y =h^{\rm DL} s + \nu
\end{equation}
where $h^{\rm DL}$ denotes the channel coefficient of the direct link. It should be noted that the performance of DL-FSO system with various atmospheric turbulence and pointing errors has been extensively studied in the literature. However, an exact analysis  for the DL-FSO system with the  effect of random fog is not available. In \cite{rahman2021_fog}, we have developed asymptotic analysis by considering exponentiated Weibull model for atmospheric turbulence with pointing errors and random fog.

\section{Statistical Results}
In this section, we develop the PDF and CDF of the resultant channel  $h= \sum_{i=1}^{N}h_ig_i$ of  RISE-FSO system. First, we find density and distribution functions of  the combined channel $ h_i=h_i^{(f)} h_i^{(t)} h_i^{(p)}$   by unifying the PDF of various atmospheric turbulence with pointing errors as given in \eqref{eq:PDF_gg_pe}, 	\eqref{eq:PDF_Malaga_pe}, and \eqref{eq:PDF_F_pe}. Next, we develop statistical results of $Z_i=h_ig_i$  using Mellin's transform. Finally, we use the MGF of $Z_i$ to get the PDF and CDF of  $Z=\sum_{i=1}^N{Z_i}$.

The PDF of the combined channel $h_{i}$ can be computed as the product of PDFs of atmospheric turbulence $h^{(t)}$,  pointing error $h^{(p)}$ and fog $h^{(f)}$. Since the PDF $h^{(tp)}$ (i.e., the product of atmospheric turbulence $h^{(t)}$, and pointing error $h^{(p)}$) is already available in the literature, we unify the PDF $h^{(tp)}$ given in \eqref{eq:PDF_gg_pe}, 	\eqref{eq:PDF_Malaga_pe}, and \eqref{eq:PDF_F_pe}  for $\cal{GG}$,  $\cal{M}$,  and  $\cal{F}$-distribution in the following Proposition 1, and use the unified  distribution of  $h^{(tp)}$ to derive the PDF of  $h^{(tp)}h^{(f)}$ using the theory of product distribution \cite{Papoulis2001}  in Theorem 1.  

\begin{my_proposition}
	A unified PDF for the combined effect of atmospheric  turbulence  and pointing errors is given as
	\begin{equation}
		f_{h_i^{(tp)}}(x)= \psi x^{\phi-1} \sum_{l=1}^{P} \zeta_l G_{p,q}^{m,n}\left(C_lx \left|
		\begin{array}{c}
			\{a_{l,w}\}_{w=1}^{p} \\ 
			\{b_{l,w}\}_{w=1}^{q}\\
		\end{array}
		\right.\right)
		\label{eq:PDF_gen_pe}
	\end{equation}
	where parameters in \eqref{eq:PDF_gen_pe} define specific atmospheric turbulence model, as given in Table \ref{table:unified_param}.

\end{my_proposition}
\begin{IEEEproof}
	We use 	\eqref{eq:PDF_gen_pe} as a general fading model and   derive the  parameters of specific models  using \eqref{eq:PDF_gg_pe}, \eqref{eq:PDF_Malaga_pe}, and 	\eqref{eq:PDF_F_pe}, as depicted in Table  \ref{table:unified_param}.
\end{IEEEproof}

\begin{table*}[tp]
	\caption{Parameters of the unified pdf  (Proposition 1)}
	\label{table:unified_param}
	\centering
\begin{tabular}{c c}
	\hline
	Turbulence Model & Unified parameters \\ \hhline{= =} 
$\cal{GG}$ & $\psi=\frac{\alpha_{G}\beta_{{\scriptscriptstyle G}}\rho^2}{A_0\Gamma(\alpha_{{\scriptscriptstyle G}})\Gamma(\beta_{{\scriptscriptstyle G}})}$, $\phi=1$, $P=1$, $\zeta_{1}=1$, $C_{1}=\frac{\alpha_{{\scriptscriptstyle G}}\beta_{{\scriptscriptstyle G}}}{A_0}$, \\ & $\{m,n,p,q\}=\{3,0,1,3\}$, $a=\{\rho^{2}\}$, $b=\{\rho^{2}-1, \alpha_{{\scriptscriptstyle G}}-1, \beta_{{\scriptscriptstyle G}}-1\}$ \\  \hline
	$\cal{M}$ & $\psi=\frac{\rho^{2}A_{mg}}{2}$, $\phi=0$, $P=\beta_{{\scriptscriptstyle M}}$, $\zeta_{1}=b_{l}$, $C_{1}=\frac{\alpha_{{\scriptscriptstyle M}} \beta_{{\scriptscriptstyle M}}}{(g\beta_{{\scriptscriptstyle M}}+\Omega^{'})A_{0}}$, \\ &  $\{m,n,p,q\}=\{3,0,1,3\}$, $a=\{\rho^{2}+1\}$, $b=\{\rho^{2}, \alpha_{{\scriptscriptstyle M}}, l\}$ \\ \hline
	$\cal{F}$ & $\psi=\frac{\alpha_{{\scriptscriptstyle F}}\rho^2}{(\beta_{{\scriptscriptstyle F}}-1)h_lA_0\Gamma(\alpha_{{\scriptscriptstyle F}})\Gamma(\beta_{{\scriptscriptstyle F}})}$, $\phi=1$, $P=1$, $\zeta_{1}=1$, $C_{1}=\frac{\alpha_{{\scriptscriptstyle F}}}{(\beta_{{\scriptscriptstyle F}}-1)h_lA_0}$, \\ &  $\{m,n,p,q\}=\{2,1,2,2\}$, $a=\{-\beta_{{\scriptscriptstyle F}},\rho^{2}$\},$b=\{\alpha_{{\scriptscriptstyle F}}-1, \rho^{2}-1\}$ \\ \hline\hline 
\end{tabular}
\end{table*}

 It is straightforward to use \eqref{eq:PDF_gen_pe} and find the PDF of combined channel $h_i= h_i^{(tp)} h_i^{(f)}$  if the channel gain $h_i^{(f)}$  is considered to be deterministic.  In the following Theorem, we develop a novel PDF considering the channel gain $h_i^{(f)}$ to be distributed according to  	\eqref{eq:h_fg} in the presence of fog.

\begin{my_theorem}
If $k$ and $v$ are the parameters of the foggy channel and Table  \ref{table:unified_param} depicts the parameters for atmospheric turbulence and pointing errors, then the PDF and CDF of the combined fading channel with atmospheric turbulence, pointing errors, and random fog    are  given by
	\begin{flalign}\label{eq:pdf_combined}
	&f_{h_i}(x) = \psi v^k x^{\phi-1} \sum_{l=1}^{P} \zeta_l  \nonumber \\& G_{p+k,q+k}^{m+k,n}\left[C_{l}x \left|\begin{array}{c}
			\{a_{l,w}\}_{w=1}^{p},\{v-\phi+1\}_{1}^{k} \\
			\{b_{l,w}\}_{w=1}^{m},\{v-\phi\}_{1}^{k},\{b_{l,w}\}_{w=m+1}^{q}
		\end{array}\right.\right]
	\end{flalign}
	
	\begin{flalign}\label{eq:cdf_combined}
		&F_{h_i}(x) = \psi v^k x^{\phi} \sum_{l=1}^{P} \zeta_l G_{p+k+1,q+k+1}^{m+k,n+1}\nonumber \\ & \left[C_{l} x\left|\begin{array}{c}
			\{a_{l,w}\}_{w=1}^{n},\{1-\phi\},\{a_{l,w}\}_{w=n+1}^{p},\{v-\phi+1\}_{1}^{k} \\
			\{b_{l,w}\}_{w=1}^{m},\{v-\phi\}_{1}^{k},\{b_{l,w}\}_{w=m+1}^{q},\{-\phi\}\\
		\end{array}\right.\right]
	\end{flalign}

\end{my_theorem}
\begin{IEEEproof} 
See Appendix A.	
\end{IEEEproof}	
Note that the parameter $k$ should be a positive integer to satisfy the definition of Meijer's G-function.    Since the underlying PDFs in 	\eqref{eq:PDF_gen_pe} and \eqref{eq:pdf_combined}  have a similar structure, the  unified performance analysis presented in this paper is applicable for both deterministic and random path loss model. As such,   the  PDF represented in  \eqref{eq:pdf_combined} can be reduced to \eqref{eq:PDF_gen_pe} for the deterministic path gain by substituting  $k=0$ and limiting the argument of Meijer's G-function up to $p$ and $q$ terms.
	
 To facilitate  performance analysis for the RISE-FSO, the distribution function of  $\sum_{i=1}^{N} h_i g_i$ is required.  Considering $L$ reflecting paths in each RIS element, we derive the PDF and CDF of  the generalized system $Z = \sum_{i=1}^{N} Z_{i}$, where  $Z_{i} = \prod_{j=1}^{L}h_{i, j}$ and $h_{i, j}, j= 1, 2, \cdots L$ are i.ni.d random variable distributed according to (\ref{eq:pdf_combined}).  Note that we can use the results of Theorem 1 to analyze the performance of the DL-FSO system.

\begin{my_proposition} If $k$ and $v$ are the parameters of the foggy channel and Table  \ref{table:unified_param} depicts the parameters for atmospheric turbulence and pointing errors, then the PDF and CDF, and MGF of a product of $L$ independent but non-identical (i.ni.d) random variables  $Z_{i}=\prod_{j=1}^{L}h_{i,j}$ are given by  
	
	\begin{eqnarray}\label{eq:pdf_prod_ris}
		&f_{Z_{i}}(x) = \frac{1}{x} \sum_{l_{1},\cdots,l_{L}=1}^{P} \prod_{j=1}^{L} \psi_{j} v_{j}^{k_{j}} \zeta_{l_{j}} \bigg(C_{l_{j}}\bigg)^{-\phi_{j}} \nonumber \\ & G_{pL+\sum_{j=1}^{L}k_{j},qL+\sum_{j=1}^{L}k_{j}}^{mL+\sum_{j=1}^{L}k_{j},nL} \left[x\prod_{j=1}^{L}C_{l_{j}}\left|\begin{array}{c}
			V_{1},V_{2} \\
			V_{3}\\
		\end{array}\right.\right]	
	\end{eqnarray}
	\begin{eqnarray}\label{eq:cdf_prod_ris}
		&F_{Z_{i}}(x) =  \sum_{l_{1},\cdots,l_{L}=1}^{P} \prod_{j=1}^{L} \psi_{j} v_{j}^{k_{j}} \zeta_{l_{j}} \bigg(C_{l_{j}}\bigg)^{-\phi_{j}} \nonumber \\ & G_{pL+\sum_{j=1}^{L}k_{j}+1,qL+\sum_{j=1}^{L}k_{j}+1}^{mL+\sum_{j=1}^{L}k_{j},nL+1} \left[x\prod_{j=1}^{L}C_{l_{j}}\left|\begin{array}{c}
			V_{1},\{1\},V_{2} \\
			V_{3},\{0\}\\
		\end{array}\right.\right]		
	\end{eqnarray}
	\begin{eqnarray}\label{eq:mgf_prod_ris}
		&M_{Z_i}(s) =  \sum_{l_{1},\cdots,l_{L}=1}^{P} \prod_{j=1}^{L} \psi_{j} v_{j}^{k_{j}} \zeta_{l_{j}} \bigg(C_{l_{j}}\bigg)^{-\phi_{j}}\nonumber \\ & G_{pL+\sum_{j=2}^{L}k_{j}+1,qL+\sum_{j=1}^{L}k_{j}}^{mL+\sum_{j=1}^{L}k_{j},nL+1} \left[\frac{1}{s}\prod_{j=1}^{L}C_{l_{j}}\left|\begin{array}{c}
			V_{1},\{1\},V_{2} \\			V_{3}\\
		\end{array}\right.\right]		
	\end{eqnarray}
\text{where} $V_{1}=\{\{\phi_{j}+a_{l_{j},w}\}_{j=1}^{L}\}_{w=1}^{n}$, $V_{2}=\{\{\phi_{j}+a_{l_{j},w}\}_{j=1}^{L}\}_{w=n+1}^{p},\{\{v_{j}+1\}_{1}^{k_{j}}\}_{j=1}^{L}$ \text{and} $V_{3}=\{\{\phi_{j}+b_{l_{j},w}\}_{j=1}^{L}\}_{w=1}^{m},\{\{v_{j}\}_{1}^{k_{j}}\}_{j=1}^{L},\{\{\phi_{j}+b_{l_{j},w}\}_{j=1}^{L}\}_{w=m+1}^{q}$.
\end{my_proposition}
\begin{IEEEproof}
	See Appendix B.
\end{IEEEproof}

\begin{my_theorem} If $k$ and $v$ are the parameters of the foggy channel and Table  \ref{table:unified_param} depicts the parameters for atmospheric turbulence and pointing errors, then the PDF and CDF of $Z=\sum_{i=1}^N{Z_i}$ are given by \eqref{eq:PDF_final} and \eqref{eq:CDF_final} respectively.
\begin{figure*}	
	\begin{eqnarray}\label{eq:PDF_final}
		f_{Z}(x) = \frac{1}{x} \sum_{l_{1,1},\cdots,l_{1,L}=1}^{P}\cdots\sum_{l_{N,1},\cdots,l_{N,L}=1}^{P}\prod_{i=1}^{N} \prod_{j=1}^{L} \psi_{i,j} v_{i,j}^{k_{i,j}} \zeta_{l_{i,j}} \bigg(C_{l_{i,j}}\bigg)^{-\phi_{i,j}}\nonumber\\ H_{0,1:pL+\sum_{j=1}^{L}k_{1,j}+1,qL+\sum_{j=1}^{L}k_{1,j};\cdots;pL+\sum_{j=1}^{L}k_{N,j}+1,qL+\sum_{j=1}^{L}k_{N,j}}^{0,0:mL+\sum_{j=1}^{L}k_{1,j},nL+1;\cdots;mL+\sum_{j=1}^{L}k_{N,j},nL+1}\left[\begin{array}{c}x \prod_{j=1}^{L} C_{l_{i,j}}\\.\\.\\.\\x \prod_{j=1}^{L} C_{l_{i,j}}\end{array} \middle\vert \begin{array}{c}
			-:V_{1} \\
			(1;1,\cdots,1):V_{2}
		\end{array}\right]			
	\end{eqnarray}
	
	\begin{eqnarray}\label{eq:CDF_final}
		F_{Z}(x) = \sum_{l_{1,1},\cdots,l_{1,L}=1}^{P}\cdots\sum_{l_{N,1},\cdots,l_{N,L}=1}^{P}\prod_{i=1}^{N} \prod_{j=1}^{L} \psi_{i,j} v_{i,j}^{k_{i,j}} \zeta_{l_{i,j}} \bigg(C_{l_{i,j}}\bigg)^{-\phi_{i,j}}\nonumber\\ H_{0,1:pL+\sum_{j=1}^{L}k_{1,j}+1,qL+\sum_{j=1}^{L}k_{1,j};\cdots;pL+\sum_{j=1}^{L}k_{N,j}+1,qL+\sum_{j=1}^{L}k_{N,j}}^{0,0:mL+\sum_{j=1}^{L}k_{1,j},nL+1;\cdots;mL+\sum_{j=1}^{L}k_{N,j},nL+1}\left[\begin{array}{c}x \prod_{j=1}^{L} C_{l_{i,j}}\\.\\.\\.\\x \prod_{j=1}^{L} C_{l_{i,j}}\end{array} \middle\vert \begin{array}{c}
			-:V_{1} \\
			(0;1,\cdots,1):V_{2}
		\end{array}\right]			
	\end{eqnarray}
	\text{where} $V_{1}=\{\{\{(\phi_{i,j}+a_{l_{i,j},w},1)\}_{j=1}^{L}\}_{w=1}^{n},(1,1),\{\{(\phi_{i,j}+a_{l_{i,j},w},1)\}_{j=1}^{L}\}_{w=n+1}^{p},\{\{(v_{i,j}+1,1)\}_{1}^{k_{i,j}}\}_{j=1}^{L}\}_{i=1}^{N}$ and $V_{2}=\{\{\{(\phi_{i,j}+b_{l_{i,j},w},1)\}_{j=1}^{L}\}_{w=1}^{m},\{\{(v_{i,j},1)\}_{1}^{k_{i,j}}\}_{j=1}^{L},\{\{(\phi_{i,j}+b_{l_{i,j},w},1)\}_{j=1}^{L}\}_{w=m+1}^{q}\}_{i=1}^{N}$.
	\hrule
\end{figure*}	
\end{my_theorem}
\begin{IEEEproof}
	See Appendix C.
\end{IEEEproof}

As a special case to simplify the notation of multi-variate Fox's H-function, we use $L=2$ in \eqref{eq:PDF_final} (which corresponds to the system model, as shown in Fig. \ref{fig:system_model}) with  $N=2$ RIS elements and similar foggy conditions depicted by $k_{i,j}=k\forall i,j$ to express the PDF  \eqref{eq:PDF_final} in terms of simpler Bi-variate Fox's H-function:
			
\small
		\begin{eqnarray}\label{eq:PDF_final_sim_iid}
			f_{Z}(x) = \frac{1}{x} \sum_{l_{1,1},l_{1,2}=1}^{P}\sum_{l_{2,1},l_{2,2}=1}^{P}\prod_{i=1}^{2} \prod_{j=1}^{2} \psi_{i,j} v_{i,j}^{k} \zeta_{l_{i,j}} \bigg(C_{l_{i,j}}\bigg)^{-\phi_{i,j}}\nonumber\\ H_{0,1:2p+2k+1,2q+2k;2p+2k+1,2q+2k}^{0,0:2m+2k,2n+1;2m+2k,2n+1}\left[\begin{array}{c}x \prod_{j=1}^{2} C_{l_{1,j}}\\x \prod_{j=1}^{2} C_{l_{2,j}}\end{array} \middle\vert \begin{array}{c}
				-:V_{1} \\
				(1;1,\cdots,1):V_{2}
			\end{array}\right]			
		\end{eqnarray}
\normalsize
	\text{where} $V_{1}=\{\{\{(\phi_{i,j}+a_{l_{i,j},w},1)\}_{j=1}^{2}\}_{w=1}^{n},(1,1),\{\{(\phi_{i,j}+a_{l_{i,j},w},1)\}_{j=1}^{2}\}_{w=n+1}^{p},\{\{(v_{i,j}+1,1)\}_{1}^{k}\}_{j=1}^{2}\}_{i=1}^{2}$ and $V_{2}=\{\{\{(\phi_{i,j}+b_{l_{i,j},w},1)\}_{j=1}^{2}\}_{w=1}^{m},\{\{(v_{i,j},1)\}_{1}^{k}\}_{j=1}^{2},\{\{(\phi_{i,j}+b_{l_{i,j},w},1)\}_{j=1}^{2}\}_{w=m+1}^{q}\}_{i=1}^{2}$.
		
Similarly, we can simplify the CDF in  \eqref{eq:CDF_final} for $L=2$, $N=2$, and $k_{i,j}=k\forall i,j$.  
In what follows, we analyze the performance of RISE-FSO and DL-FSO systems using the statistical results of Theorem 1 and Theorem 2, respectively.
\section{Performance Analysis}
In this section, we analyze the performance of the RISE-FSO by a simple customization  of Theorem 2 with  $L=2$, $z_{i,1}=h_i$, and $z_{i, 2}=g_i$, which corresponds to the RISE-FSO system model in \eqref{model_smpl}. Thus, we denote the resultant by $h= \sum_{i=1}^Nh_i g_i$.  Finally, we introduce the third unification of performance evaluation with HD ($t=1$) and IM/DD ($t=2$) detection modes by defining the SNR of the system as $\gamma=\gamma_0 h^t$, where  $\gamma_0= \frac{P_T^t}{\sigma_\nu^2}$. Using a straightforward transformation of the random variable, the CDF of SNR $F_{\gamma}(\gamma)={\rm Pr}(\gamma_{0}h^{t}\le\gamma) = {\rm Pr}(h\le(\frac{\gamma}{\gamma_{0}})^{1/t}) = F_{h}((\frac{\gamma}{\gamma_{0}})^{1/t})$ is given by
\begin{equation}\label{eq:cdf_snr_ris}
	F_{\gamma_{}}(\gamma) = F_{h}\bigg(\bigg(\frac{\gamma}{\gamma_0}\bigg)^{1/t}\bigg)
\end{equation}
where $F_h(\cdot)$ is given in \eqref{eq:CDF_final} with $L=2$. Using \eqref{eq:cdf_snr_ris}, the PDF of SNR can be expressed as
\begin{equation}\label{eq:pdf_snr_ris}
	f_{\gamma_{}}(\gamma) = \frac{1}{t \gamma_0^{\frac{1}{t}} \gamma^{1-\frac{1}{t}}} f_{h}\bigg(\bigg(\frac{\gamma}{\gamma_0}\bigg)^{1/t}\bigg) 
\end{equation}
where $f_h(\cdot)$ is given in \eqref{eq:PDF_final} with $L=2$. Similarly, we use  \eqref{eq:pdf_combined} and \eqref{eq:cdf_combined}   to express the PDF and CDF of SNR for the DL-FSO  as

\begin{eqnarray}
	\label{eq:cdfpdf_zaf}
f_{\gamma_{}}^{\rm DL}(\gamma) = \frac{1}{t \gamma_0^{\frac{1}{t}} \gamma^{1-\frac{1}{t}}} f_{h_{i}}\bigg(\big(\frac{\gamma}{\gamma_0}\big)^{1/t}\bigg),\nonumber\\ F_{\gamma_{}}^{\rm DL }(\gamma) = F_{h_{i}}\bigg(\big(\frac{\gamma}{\gamma_0}\big)^{1/t}\bigg)
\end{eqnarray}
Using $t=1$ and $t=2$ in \eqref{eq:cdfpdf_zaf}, the PDF and CDF of the SNR for HD and IM/DD detection techniques can be verified for  the $\mathcal{GG}$  turbulence (see \cite{Long2020} and references therein).

\subsection{Outage Probability}\label{sec:op}
Outage probability is a performance metric to characterize the impact of fading in a communication system. Mathematically, it can be defined as  the probability of SNR falling below a threshold value $\gamma_{\rm th}$ i.e.,  $ P_{\rm out}=Pr(\gamma \le \gamma_{\rm th})$.

\begin{my_lemma}\label{lemma_out_zaf}
	We present  the following results of outage probability for the RISE-FSO system:
	\begin{enumerate}[label=(\alph*)]
		\item  An exact expression for the outage probability is given as  $P_{\rm out} = F_{\gamma}(\gamma_{\rm th})$. 
		\item Asymptotically at a high SNR,  the outage probability is given by \eqref{eq:out_high_ris}.

	\item The diversity order is given as $G_{\rm out} = \sum_{i=1}^{N} \min\{\{\frac{\phi_{i,1}+b_{l_{i,1},w}}{t}\}_{w=1}^{m},\frac{v_{i,1}}{t}, \{\frac{\phi_{i,2}+b_{l_{i,2},w}}{t}\}_{w=1}^{m},\frac{v_{i,2}}{t}\}$.
	\end{enumerate}
	
\end{my_lemma}
\begin{IEEEproof}
Part (a) can be obtained  using the direct definition of the outage probability. To prove part (b), we use \cite[Eqn. (30)]{Rahama2018} to derive the outage probability asymptotically at a high SNR $\gamma_{0}\rightarrow\infty$ in terms of Gamma function.  As such,  the asymptotic expression in \eqref{eq:out_high_ris}  is obtained by computing the residue of multiple Mellin-Barnes integrals of the corresponding multi-variate Fox's H-function at the dominant pole $p_{i} = \min\{\{\phi_{i,j}+b_{l_{i,j},w}\}_{w=1}^{m},v_{i,j}\}_{j=1}^{2}$. To prove (c), we need to express \eqref{eq:out_high_ris} as $P_{out}^{\infty} \propto \gamma_{0}^{-G_{\rm out}}$ in order to get the outage-diversity order $G_{\rm out}$ of the system. Using the dominant pole $p_i$, $\forall l_{i,j},j$ (i.e., considering over all the summation terms in \eqref{eq:out_high_ris}), the $N$-products of the term $\gamma_0^{-\frac{p_i}{t}}$  result into $P_{\rm out}^{\infty} \propto \gamma_{0}^{-\sum_{i}^{N}\frac{p_{i}}{t}}$. Finally, we use  $p_{i}$ to get the diversity order $ G_{\rm out}$ as given in part (c).
\end{IEEEproof}

\small	
\begin{figure*}
		\begin{eqnarray}\label{eq:out_high_ris}
		P_{\rm out}^{\infty} \approx \sum_{l_{1,1},l_{1,2}=1}^{P}\cdots\sum_{l_{N,1},l_{N,2}=1}^{P}\prod_{i=1}^{N} \prod_{j=1}^{2} \psi_{i,j} v_{i,j}^{k_{i,j}} \zeta_{l_{i,j}} \big(C_{l_{i,j}}\big)^{-\phi_{i,j}} \bigg(\frac{\gamma_{th}}{\gamma_0}\bigg)^{p_{i}/t} \bigg(\prod_{j=1}^{2} C_{l_{i,j}}\bigg)^{p_{i}} \frac{1}{\Gamma(1+\sum_{i=1}^{N}p_{i})}  \nonumber \\ \prod_{i=1}^{N}  \frac{\prod_{j=1}^{2}\prod_{w=1,p_{i}\neq \phi_{i,j}+b_{l_{i,j},w}}^{m}\Gamma(\phi_{i,j}+b_{l_{i,j},w}-p_{i})\prod_{j=1,p_{i}\neq v_{i,j}}^{2}\big(\Gamma(v_{i,j}-p_{i})\big)^{k_{i,j}} \prod_{j=1}^{2} \prod_{w=1}^{n}\Gamma(1-\phi_{i,j}-a_{l_{i,j},w}+p_{i}) \Gamma(p_{i})}{\prod_{j=1}^{2} \prod_{w=n+1}^{p}\Gamma(\phi_{i,j}+a_{l_{i,j},w}-p_{i})\prod_{j=1}^{2}\big(\Gamma(v_{i,j}+1-p_{i})\big)^{k_{i.j}}\prod_{j=1}^{2}\prod_{w=m+1}^{q}\Gamma(1-\phi_{i,j}-b_{l_{i,j},w}+p_{i})} 	
	\end{eqnarray}
	where $p_{i} = \min\{\{\phi_{i,j}+b_{l_{i,j},w}\}_{w=1}^{m},v_{i,j}\}_{j=1}^{2}$.
	\hrule
\end{figure*}
\normalsize

\small
\begin{figure*}
	\begin{eqnarray}\label{eq:out_high}
		P_{\rm {out}}^{\infty, \rm DL } \approx \psi v^k \sum_{l=1}^{P} \zeta_l   C_{l}^{-\phi} \Bigg[\sum_{i=1}^{m}\frac{\prod_{w=1,w\neq i}^{m}\Gamma(b_{l,w}-b_{l,i})\big(\Gamma(v-\phi-b_{l,i})\big)^{k}\prod_{w=1}^{n}\Gamma(1-a_{l,w}+b_{l,i})\Gamma(\phi+b_{l,i})}{\prod_{w=n+1}^{p}\Gamma(a_{l,w}-b_{l,i})\big(\Gamma(v+1-\phi-b_{l,i})\big)^{k}\prod_{w=m+1}^{q}\Gamma(1-b_{l,w}+b_{l,i})\Gamma(1+\phi+b_{l,i})} \nonumber \\
		\bigg(C_{l}^{\phi+b_{l,i}} \bigg(\frac{\gamma_{th}}{\gamma_0}\bigg)^{(\phi+b_{l,i})/t}\bigg) + k \frac{\prod_{w=1}^{m}\Gamma(\phi+b_{l,w}-v)\prod_{w=1}^{n}\Gamma(1-\phi-a_{l,w}+v)\Gamma(v)}{\prod_{w=n+1}^{p}\Gamma(\phi+a_{l,w}-v)\prod_{w=m+1}^{q}\Gamma(1-\phi-b_{l,w}+v)\Gamma(1+v)} \bigg(C_{l}^{v} \bigg(\frac{\gamma_{th}}{\gamma_0}\bigg)^{v/t}\bigg)\Bigg]
	\end{eqnarray}
	\hrule	
\end{figure*}
\normalsize

We present  Table \ref{table:diversity_order} with $\kappa=0$  for the diversity order of RISE-FSO system for three turbulence models and both the HD and IM/DD detection modes. It can be seen that the outage performance of the RISE-FSO improves with an increase in the number of RIS elements.

For the DL-FSO system, an exact expression of the outage probability  can be obtained using the CDF in \eqref{eq:cdfpdf_zaf} and  \eqref{eq:cdf_combined} as   $P^{\rm DL}_{\rm out} = F^{\rm DL}_{\gamma}(\gamma_{\rm th})$.  We use  the asymptotic representation of the Meijer's G-function \cite[07.34.06.0005.01]{Meijers} on the derived $P^{\rm DL}_{\rm out}$ to express the outage probability in the high SNR regime $\gamma_{0} \rightarrow \infty$, as \eqref{eq:out_high}.

To derive the diversity order of the DL-FSO system, we require the dominant term  of  $\gamma_0^{-(\phi+b_{l,i})/t}$ and  $\gamma_0^{-(v)/t}$ over $P$ and $m$ summation terms in \eqref{eq:out_high}. Thus, we express $P_{\rm {out}}^{\infty, \rm DL } \propto \gamma_{0}^{-G_{\rm out}^{\rm DL}}$  using the minimum exponent of $\gamma_0$ to get $G_{\rm out}^{\rm DL} = \min(\{\frac{\phi+b_{l,i}}{t}\}_{l=1,i=1}^{l=P, i=m},\frac{v}{t})$. Using Table \ref{table:unified_param}, the diversity order for $\cal{F}$, $\cal{GG}$, and $\cal{M}$ turbulence with pointing errors and random fog parameters are respectively, $\min\{\frac{\alpha_{{\scriptscriptstyle F}}}{t},\frac{\rho^{2}}{t},\frac{v}{t}\}$,  $ \min\{\frac{\rho^{2}}{t},\frac{\alpha_{{\scriptscriptstyle G}}}{t},\frac{\beta_{{\scriptscriptstyle G}}}{t},\frac{v}{t}\}$, and $ \min\{\frac{\rho^{2}}{t},\frac{\alpha_{{\scriptscriptstyle M}}}{t},\frac{\beta_{{\scriptscriptstyle M}}}{t},\frac{v}{t}\}$. This is consistent with previous results of diversity order for DL-FSO systems with atmospheric turbulence and pointing errors with deterministic path loss (see \cite{Ansari2016} \cite{dual_hop_turb2017} \cite{Badarneh2020}, and references therein).

\begin{table*}[tp]
	\caption{Diversity order of RISE-FSO }
	\label{table:diversity_order}
	\centering
	\begin{tabular}{c c}
		\hline
		Turbulence Model  & $G_{\rm out}$ with $\kappa=0$ and $G_{\rm BER}$ with $\kappa=1$ \\ \hhline{= =} 
		$\cal{GG}$ & $\sum_{i=1}^{N} \min\{\frac{\alpha_{{\scriptscriptstyle G}}(i,1)-\kappa}{t},\frac{\beta_{{\scriptscriptstyle G}}(i,1)-\kappa}{t},\frac{\rho^{2}_{i,1}-\kappa}{t},\frac{v_{i,1}-\kappa}{t},\frac{\alpha_{{\scriptscriptstyle G}}(i,2)-\kappa}{t},\frac{\beta_{{\scriptscriptstyle G}}(i,2)-\kappa}{t},\frac{\rho^{2}_{i,2}-\kappa}{t},\frac{v_{i,2}-\kappa}{t}\}$ \\  \hline
		$\cal{M}$ & $\sum_{i=1}^{N} \min\{\frac{\alpha_{{\scriptscriptstyle M}}(i,1)-\kappa}{t},\frac{\beta_{{\scriptscriptstyle M}}(i,1)-\kappa}{t},\frac{\rho^{2}_{i,1}-\kappa}{t},\frac{v_{i,1}-\kappa}{t},\frac{\alpha_{{\scriptscriptstyle M}}(i,2)-\kappa}{t},\frac{\beta_{{\scriptscriptstyle M}}(i,2)-\kappa}{t},\frac{\rho^{2}_{i,2}-\kappa}{t},\frac{v_{i,2}-\kappa}{t}\}$ \\ \hline
		$\cal{F}$ & $\sum_{i=1}^{N} \min\{\frac{\alpha_{{\scriptscriptstyle F}}(i,1)-\kappa}{t},\frac{\rho^{2}_{i,1}-\kappa}{t},\frac{v_{i,1}-\kappa}{t},\frac{\alpha_{{\scriptscriptstyle F}}(i,2)-\kappa}{t},\frac{\rho^{2}_{i,2}-\kappa}{t},\frac{v_{i,2}-\kappa}{t}\}$ \\ \hline \hline
		\end{tabular}
\end{table*}

\subsection{ Average BER}\label{sec:ber}
Average BER is used to quantify the reliability of data transmissions. For binary modulations, the average BER using the CDF of SNR is given as \cite{Ansari2011_ber}
\begin{equation}\label{eq:gen_ber_fso}
	\bar{P}_{e} =  \frac{q^{p}}{2\Gamma(p)} \int_{0}^{\infty} \gamma^{p-1} e^{-q\gamma} F_{\gamma} (\gamma) d\gamma
\end{equation}
where $p$ and $q$ are modulation specific parameters. Specifically, for
coherent binary FSK (CBFSK), $p = q = 0.5$, coherent binary PSK (CBPSK), $p = 0.5$ and $q = 1$, non-coherent binary FSK (NBFSK), $p = 1$ and $q = 0.5$, differential binary PSK  (DBPSK), $p = 1$ and $q = 1$. 
\begin{my_lemma} We present the following results of the average BER for the RISE-FSO system:
	\begin{enumerate}[label=(\alph*)]
	\item  An exact expression for the average BER is given by \eqref{eq:ber_ris_fso}.
\small
\begin{figure*}
\begin{eqnarray}\label{eq:ber_ris_fso}
	\bar{P}_{e} = \bigg(\frac{1}{2\Gamma(p)}\bigg) \sum_{l_{1,1},l_{1,2}=1}^{P}\cdots\sum_{l_{N,1},l_{N,2}=1}^{P}\prod_{i=1}^{N} \prod_{j=1}^{2} \psi_{i,j} v_{i,j}^{k_{i,j}} \zeta_{l_{i,j}} \bigg(C_{l_{i,j}}\bigg)^{-\phi_{i,j}} \nonumber \\H_{1,1:2p+k_{1,1}+k_{1,2}+1,2q+k_{1,1}+k_{1,2};\cdots;2p+k_{N,1}+k_{N,2}+1,2q+k_{N,1}+k_{N,2}}^{0,1:2m+k_{1,1}+k_{1,2},2n+1;\cdots;2m+k_{N,1}+k_{N,2},2n+1}  \left[\begin{array}{c}\bigg(\frac{1}{q\gamma_0}\bigg)^{1/t} \prod_{j=1}^{2} C_{l_{i,j}}\\.\\.\\.\\\bigg(\frac{1}{q\gamma_0}\bigg)^{1/t} \prod_{j=1}^{2} C_{l_{i,j}}\end{array} \middle\vert \begin{array}{c}
		(1-p,\frac{1}{t},\cdots,\frac{1}{t}):V_{1} \\
		(0;1,\cdots,1):V_{2}
	\end{array}\right]			
\end{eqnarray}
\normalsize
\text{where} $V_{1}=\{\{(\phi_{i,1}+a_{l_{i,1},w},1)\}_{w=1}^{n},\{(\phi_{i,2}+a_{l_{i,2},w},1)\}_{w=1}^{n},(1,1),\{(\phi_{i,1}+a_{l_{i,1},w},1)\}_{w=n+1}^{p},\{(\phi_{i,2}+a_{l_{i,2},w},1)\}_{w=n+1}^{p},\{(v_{i,1}+1,1)\}_{1}^{k_{i,1}},\{(v_{i,2}+1,1)\}_{1}^{k_{i,2}}\}_{i=1}^{N}$ \text{and} $V_{2}=\{\{(\phi_{i,1}+b_{l_{i,2},w},1)\}_{w=1}^{m},\{(\phi_{i,2}+b_{l_{i,2},w},1)\}_{w=1}^{m},\{(v_{i,1},1)\}_{1}^{k_{i,1}},\{(v_{i,2},1)\}_{1}^{k_{i,2}},\{(\phi_{i,1}+b_{l_{i,1},w},1)\}_{w=m+1}^{q},\{(\phi_{i,2}+b_{l_{i,2},w},1)\}_{w=m+1}^{q}\}_{i=1}^{N}$.
    \hrule	
\end{figure*}
	\item Asymptotically at high SNR $\gamma_{0}\rightarrow\infty$, the average BER can be expressed as \eqref{eq:ber_high_ris_fso}.
\small
\begin{figure*}
	\begin{eqnarray}\label{eq:ber_high_ris_fso}
		\bar{P}_{e}^{\infty} \approx \sum_{l_{1,1},l_{1,2}=1}^{P}\cdots\sum_{l_{N,1},l_{N,2}=1}^{P}\prod_{i=1}^{N} \prod_{j=1}^{2} \frac{\psi_{i,j}v_{i,j}^{k_{i,j}}}{2\Gamma(p)}  \zeta_{l_{i,j}} \big(C_{l_{i,j}}\big)^{N+p_{i}-1-\phi_{i,j}} \bigg(\frac{1}{q\gamma_0}\bigg)^{\frac{p_{i}-1}{t}} \bigg(\frac{1}{q}\bigg)^{N/t}  \frac{\Gamma(p+\frac{1}{t}\sum_{i=1}^{N}p_{i})}{\Gamma(1+\sum_{i=1}^{N}p_{i})} \nonumber \\ \prod_{i=1}^{N} \frac{\prod_{j=1}^{2}\prod_{w=1,p_{i}\neq \phi_{i,j}+b_{l_{i,j},w}}^{m}\Gamma(\phi_{i,j}+b_{l_{i,j},w}-p_{i})\prod_{j=1,p_{i}\neq v_{i,j}}^{2}\big(\Gamma(v_{i,j}-p_{i})\big)^{k_{i,j}} \prod_{j=1}^{2}\prod_{w=1}^{n}\Gamma(1-\phi_{i,j}-a_{l_{i,j},w}+p_{i}) \Gamma(p_{i})}{\prod_{j=1}^{2}\prod_{w=n+1}^{p}\Gamma(\phi_{i,j}+a_{l_{i,j},w}-p_{i})\prod_{j=1}^{2}\big(\Gamma(v_{i,j}+1-p_{i})\big)^{k_{i.j}}\prod_{j=1}^{2}\prod_{w=m+1}^{q}\Gamma(1-\phi_{i,j}-b_{l_{i,j},w}+p_{i})} 	
	\end{eqnarray}
	where $p_{i} = \min\{\{\phi_{i,1}+b_{l_{i,1},w}\}_{w=1}^{m},v_{i,1},\{\phi_{i,2}+b_{l_{i,2},w}\}_{w=1}^{m},v_{i,2}\}$.
    \hrule
\end{figure*}
\normalsize

	\item The diversity order using the average BER is $G_{\rm BER} = \sum_{i=1}^{N}  \min\{\{\frac{\phi_{i,j}+b_{l_{i,j},w}-1}{t}\}_{w=1}^{m},\frac{v_{i,j}-1}{t}\}_{j=1}^{2}$.
\end{enumerate}

\end{my_lemma}

\begin{IEEEproof} To prove (a), we substitute the CDF of the RISE-FSO system of \eqref{eq:cdf_snr_ris} (which requires \eqref{eq:CDF_final}) in \eqref{eq:gen_ber_fso},  use the definition of multivariate Fox's H-function and  interchange the order of integration to get
\small
\begin{eqnarray}\label{eq:ber_ris_proof1}
	&\bar{P}_{e} = \frac{q^{p}}{2\Gamma(p)}  \sum_{l_{1,1},l_{1,2}=1}^{P}\cdots\sum_{l_{N,1},l_{N,2}=1}^{P}\prod_{i=1}^{N} \prod_{j=1}^{2} \psi_{i,j} v_{i,j}^{k_{i,j}} \zeta_{l_{i,j}} \nonumber \\ &\big(C_{l_{i,j}}\big)^{-\phi_{i,j}} \Bigg(\bigg(\frac{1}{2\pi \J}\bigg)^{N} \int\limits_{\mathcal{L}_{i}} \bigg(\bigg(\frac{1}{\gamma_0}\bigg)^{1/t}\prod_{j=1}^{2} C_{l_{i,j}}\bigg)^{x_{i}} \nonumber \\ &\Bigg[\frac{\prod_{j=1}^{2}\prod_{w=1}^{m} \Gamma\big(-x_{i}+\phi_{i,j}+b_{l_{i,j},w}\big) }{\prod_{j=1}^{2}\prod_{w=n+1}^{p} \Gamma\big(-x_{i}+\phi_{i,j}+a_{l_{i,j},w}\big) } \nonumber \\ &\frac{ \prod_{j=1}^{2}\prod_{w=1}^{n} \Gamma\big(1+x_{i}-\phi_{i,j}-a_{l_{i,j},w}\big)}{ \prod_{j=1}^{2}\prod_{w=m+1}^{q} \Gamma\big(1+x_{i}-\phi_{i,j}-b_{l_{i,j},w}\big)}\frac{\prod_{j=1}^{2}\big(\Gamma\big(v_{i,j}-x_{i}\big)\big)^{k_{i,j}}}{\prod_{j=1}^{2}\big(\Gamma\big(1+v_{i,j}-x_{i}\big)\big)^{k_{i,j}}} \nonumber \\ &\frac{\Gamma(x_{i})}{\Gamma\big(1+\sum_{i=1}^{N}x_{i}\big)}\Bigg]\bigg(\int_{0}^{\infty} e^{-q\gamma} \gamma^{p-1} \gamma^{\frac{1}{t}\sum_{i=1}^{N} x_{i}}  d\gamma \bigg) \diff x_i\Bigg) 			
\end{eqnarray}
\normalsize

We solve  the inner integral in \eqref{eq:ber_ris_proof1}:
\begin{equation}
	\label{eq:zaf_ber2}
	\int_{0}^{\infty} e^{-q\gamma} \big(\gamma\big)^{p-1+\frac{1}{t}\sum_{i=1}^{N} x_{i}} d\gamma = \bigg(\frac{1}{q} \bigg)^{p+\frac{1}{t}\sum_{i=1}^{N} x_{i}} \Gamma \bigg(p+\frac{1}{t}\sum_{i=1}^{N} x_i \bigg)
\end{equation}
Using \eqref{eq:zaf_ber2} in \eqref{eq:ber_ris_proof1}, we apply the definition of $N$-multivariate Fox's H-function \cite[A.1]{M-Foxh} to get \eqref{eq:ber_ris_fso}. To prove (b),  we  use the asymptotic analysis in \cite[(31)]{Rahama2018} to express the average BER at a high SNR in \eqref{eq:ber_high_ris_fso}.	To prove (c), we express $\bar{P}_{e}^{\infty} \propto \gamma_{0}^{-G_{\rm BER}}$ using the similar procedure as depicted in deriving the outage-diversity order (see proof of Lemma \ref{lemma_out_zaf}, part (c)).
\end{IEEEproof}
Similar to the outage probability, we list the diversity order of the RISE-FSO system  using the  average BER in Table \ref{table:diversity_order} with $\kappa=1$. The diversity order shows that the performance of the RISE FSO system improves with an increase in the number of RIS elements.

The average BER of the DL-FSO system can be derived using \eqref{eq:cdf_combined} in \eqref{eq:gen_ber_fso} and applying the similar procedure used in  RISE-FSO with the inner integral  $\int_{0}^{\infty} e^{-q\gamma} \gamma^{p-1} \gamma^{x/t}d\gamma 
= \frac{1}{q^{p+\frac{x}{t}}} \Gamma\big(p+\frac{x}{t}\big)$ to get \eqref{eq:ber_fso}.
\begin{figure*}
		\begin{eqnarray}\label{eq:ber_fso}
		&\hspace{-4mm}\bar{P}_{e}^{\rm DL} = \frac{\psi v^k}{2\Gamma(p)} \sum_{l=1}^{P} \zeta_l   C_{l}^{-\phi} H_{p+k+2,q+k+1}^{m+k,n+2}\left[C_{l} \bigg(\frac{1}{q\gamma_0}\bigg)^{1/t}\left|\begin{array}{c}
			\{(\phi+a_{l,w},1)\}_{w=1}^{n},(1,1),(1-p,\frac{1}{t}),\{(\phi+a_{l,w},1)\}_{w=n+1}^{p},\{(v+1,1)\}_{1}^{k} \\
			\{(\phi+b_{l,w},1)\}_{w=1}^{m},\{(v,1)\}_{1}^{k},\{(\phi+b_{l,w},1)\}_{w=m+1}^{q},(0,1)
		\end{array}\right.\right]
	\end{eqnarray}
    \hrule	
\end{figure*}
We use the asymptotic analysis of univariate Fox's H-function provided in \cite{Alhennawi2016} to express average BER of the DL-FSO at a high SNR as \eqref{eq:ber_high_fso}.

\small
\begin{figure*}
	\begin{eqnarray}\label{eq:ber_high_fso}
	\bar{P}_{e}^{\infty,\rm DL} \approx \frac{\psi v^k}{2\Gamma(p)}  \sum_{l=1}^{P} \zeta_l   C_{l}^{-\phi} \Bigg[\sum_{i=1}^{m}\frac{\prod_{w=1,w\neq i}^{m}\Gamma(b_{l,w}-b_{l,i})\big(\Gamma(v-\phi-b_{l,i})\big)^{k} \prod_{w=1}^{n}\Gamma(1-a_{l,w}+b_{l,i})\Gamma(\phi+b_{l,i})\Gamma(p+\frac{\phi+b_{l,i}}{t})}{\prod_{w=n+1}^{p}\Gamma(a_{l,w}-b_{l,i})\big(\Gamma(v+1-\phi-b_{l,i})\big)^{k}\prod_{w=m+1}^{q}\Gamma(1-b_{l,w}+b_{l,i})\Gamma(1+\phi+b_{l,i})} \nonumber \\
	\bigg(C_{l}^{\phi+b_{l,i}} 	\bigg(\frac{1}{q\gamma_0}\bigg)^{(\phi+b_{l,i})/t}\bigg) + k \frac{\prod_{w=1}^{m}\Gamma(\phi+b_{l,w}-v)\prod_{w=1}^{n}\Gamma(1-\phi-a_{l,w}+v)\Gamma(v)\Gamma(p+\frac{v}{t})}{\prod_{w=n+1}^{p}\Gamma(\phi+a_{l,w}-v)\prod_{w=m+1}^{q}\Gamma(1-\phi-b_{l,w}+v)\Gamma(1+v)} \bigg(C_{l}^{v} \bigg(\frac{1}{q\gamma_0}\bigg)^{v/t}\bigg)\Bigg]
\end{eqnarray}
    \hrule
\end{figure*}
\normalsize
To derive the BER-diversity order of the DL-FSO system, we express  $\bar{P}_{e}^{\infty,\rm DL} \propto \gamma_{0}^{-G_{\rm BER}^{\rm DL}}$ using the minimum exponent of  $\gamma_0^{-(\phi+b_{l,i})/t}$ and  $\gamma_0^{-(v)/t}$ over $P$ and $m$ summation terms in \eqref{eq:ber_high_fso} to get $G_{\rm BER}^{\rm DL} = \min(\{\frac{\phi+b_{l,i}}{t}\}_{l=1,i=1}^{l=P,i=m},\frac{v}{t})$.  Hence, the diversity order of the DL-FSO system for $\cal{F}$, $\cal{GG}$, and $\cal{M}$ are given, respectively as $\min\{\frac{\alpha_{{\scriptscriptstyle F}}}{t},\frac{\rho^{2}}{t},\frac{v}{t}\}$, $ \min\{\frac{\rho^{2}}{t},\frac{\alpha_{{\scriptscriptstyle G}}}{t},\frac{\beta_{{\scriptscriptstyle G}}}{t},\frac{v}{t}\}$, and $ \min\{\frac{\rho^{2}}{t},\frac{\alpha_{{\scriptscriptstyle M}}}{t},\frac{\beta_{{\scriptscriptstyle M}}}{t},\frac{v}{t}\}$.

\subsection{Ergodic Capacity}\label{sec:capacity}
Assuming a Gaussian codebook at the channel input, the ergodic capacity of an FSO system  defined as the maximum information transmission rate with an arbitrarily low error probability is given as \cite{Khalighi2014}:
\begin{equation}\label{eq:gen_cap_fso}
	\bar{\eta} =  \Aver{\log_2 (1 + \mu_{t} \gamma)} 
	= \int_{0}^{\infty} \log_2 (1 + \mu_{t} \gamma)  f_{\gamma} (\gamma) d\gamma 
\end{equation}
where $t \in \{1,2\}$ with $\mu_{1}=1$ HD and $\mu_{2}=\frac{e}{2\pi}$ for IM/DD receivers.
\begin{my_lemma}
		 The ergodic capacity of the RISE-FSO is given by \eqref{eq:cap_ris_fso}.
\small
\begin{figure*}	
\begin{eqnarray}\label{eq:cap_ris_fso}
	&\bar{\eta} = \frac{\log_2(e)}{t}  \sum_{l_{1,1},l_{1,2}=1}^{P}\cdots\sum_{l_{N,1},l_{N,2}=1}^{P}\prod_{i=1}^{N} \prod_{j=1}^{2} \psi_{i,j} v_{i,j}^{k_{i,j}} \zeta_{l_{i,j}} \bigg(C_{l_{i,j}}\bigg)^{-\phi_{i,j}}\nonumber\\ &H_{1,1:2p+k_{1,1}+k_{1,2}+1,2q+k_{1,1}+k_{1,2};\cdots;2p+k_{N,1}+k_{N,2}+1,2q+k_{N,1}+k_{N,2};2,2}^{0,1:2m+k_{1,1}+k_{1,2},2n+1;\cdots;2m+k_{N,1}+k_{N,2},2n+1;1,2}\left[\begin{array}{c}\bigg(\frac{1}{\epsilon \gamma_0}\bigg)^{1/t} \prod_{j=1}^{2} C_{l_{i,j}}\\.\\.\\.\\\bigg(\frac{1}{\epsilon \gamma_0}\bigg)^{1/t} \prod_{j=1}^{2} C_{l_{i,j}}\\\frac{\mu_{t}}{\epsilon}\end{array} \middle\vert \begin{array}{c}
		(1,\frac{1}{t},\cdots,\frac{1}{t},1):V_{1};(1,1),(1,1) \\
		(1;1,\cdots,1,0):V_{2};(1,1),(0,1)
	\end{array}\right]
\end{eqnarray}

\normalsize

\text{where} $V_{1}=\{\{(\phi_{i,1}+a_{l_{i,1},w},1)\}_{w=1}^{n},\{(\phi_{i,2}+a_{l_{i,2},w},1)\}_{w=1}^{n},(1,1),\{(\phi_{i,1}+a_{l_{i,1},w},1)\}_{w=n+1}^{p},\{(\phi_{i,2}+a_{l_{i,2},w},1)\}_{w=n+1}^{p},\{(v_{i,1}+1,1)\}_{1}^{k_{i,1}},\{(v_{i,2}+1,1)\}_{1}^{k_{i,2}}\}_{i=1}^{N}$ \text{and} $V_{2}=\{\{(\phi_{i,1}+b_{l_{i,1},w},1)\}_{w=1}^{m},\{(\phi_{i,2}+b_{l_{i,2},w},1)\}_{w=1}^{m},\{(v_{i,1},1)\}_{1}^{k_{i,1}},\{(v_{i,2},1)\}_{1}^{k_{i,2}},\{(\phi_{i,1}+b_{l_{i,1},w},1)\}_{w=m+1}^{q},\{(\phi_{i,2}+b_{l_{i,2},w},1)\}_{w=m+1}^{q}\}_{i=1}^{N}$.
	\hrule
\end{figure*}
\end{my_lemma}
\begin{IEEEproof}	We substitute the PDF of SNR of RISE-FSO system \eqref{eq:pdf_snr_ris} through \eqref{eq:PDF_final} in \eqref{eq:gen_cap_fso},  use the definition of multivariate Fox's H-function, and change the order of integration to get
\small
\begin{eqnarray}\label{eq:cap_expn_fso_ris}
	&\bar{\eta} = \frac{\log_2(e)}{t} \sum_{l_{1,1},l_{1,2}=1}^{P}\cdots\sum_{l_{N,1},l_{N,2}=1}^{P}\prod_{i=1}^{N} \prod_{j=1}^{2} \psi_{i,j} v_{i,j}^{k_{i,j}} \nonumber \\ &\zeta_{l_{i,j}} \big(C_{l_{i,j}}\big)^{-\phi_{i,j}} \Bigg(\bigg(\frac{1}{2\pi \J}\bigg)^{N} \int\limits_{\mathcal{L}_{i}} \bigg(\bigg(\frac{1}{\gamma_0}\bigg)^{1/t}\prod_{j=1}^{2} C_{l_{i,j}}\bigg)^{n_{i}} \nonumber \\ &\Bigg[\frac{\prod_{j=1}^{2}\prod_{w=1}^{m} \Gamma\big(-x_{i}+\phi_{i,j}+b_{l_{i,j},w}\big) }{\prod_{j=1}^{2}\prod_{w=n+1}^{p} \Gamma\big(-x_{i}+\phi_{i,j}+a_{l_{i,j},w}\big) } \nonumber \\ &\frac{ \prod_{j=1}^{2}\prod_{w=1}^{n} \Gamma\big(1+x_{i}-\phi_{i,j}-a_{l_{i,j},w}\big)}{\prod_{j=1}^{2}\prod_{w=m+1}^{q} \Gamma\big(1+x_{i}-\phi_{i,j}-b_{l_{i,j},w}\big)} \frac{ \prod_{j=1}^{2}\big(\Gamma\big(v_{i,j}-x_{i}\big)\big)^{k_{i,j}}}{\prod_{j=1}^{2}\big(\Gamma\big(1+v_{i,j}-x_{i}\big)\big)^{k_{i,j}}} \nonumber \\ &\frac{\Gamma(x_{i})}{\Gamma\big(\sum_{i=1}^{N}x_i\big)}\Bigg]
	 \bigg(\int_{0}^{\infty} \ln (1 + \mu_{t} \gamma) \gamma^{-1+\frac{1}{t}\sum_{i=1}^{N} x_{i}}  d\gamma \bigg) \diff x_{i}\Bigg)			
\end{eqnarray}
\normalsize
To solve the inner integral in \eqref{eq:cap_expn_fso_ris}, we use \cite[01.04.07.0002.01]{Meijers} to express $\ln (1 + \mu_{t} \gamma) = \frac{1}{2\pi \J} \int_{\mathcal{L}} \frac{\Gamma(u+1)\Gamma(-u)^{2}}{\Gamma(1-u)} \big(\mu_{t}\gamma\big)^{-u} \diff u$ and use the final value theorem $\lim_{x\rightarrow \infty} \int_{0}^{x}f(u) \diff u = \lim_{s\rightarrow 0} F(s) = F(\epsilon)$, where $\epsilon$ is close to zero (in the order $10^{-6})$. Thus, the inner integral becomes

\small
\begin{eqnarray}\label{eq:cap_log_int}
	&\int_{0}^{\infty} \ln (1 + \mu_t \gamma) \big(\gamma\big)^{-1+\frac{1}{t}\sum_{i=1}^{N} x_{i}} d\gamma = \lim_{s\rightarrow 0} \frac{1}{2\pi \J} \nonumber \\&\int_{\mathcal{L}} \frac{\Gamma(u+1)\Gamma(-u)^{2}}{\Gamma(1-u)} \big(\mu_{t}^{-u} \big(\frac{1}{s} \big)^{-u+\frac{1}{t}\sum_{i=1}^{N} x_{i}} \Gamma \big(-u+\frac{1}{t}\sum_{i=1}^{N} x_{i} \big) \big) \diff u \nonumber \\ &\hspace{-8mm}= \frac{1}{2\pi \J} \int_{\mathcal{L}} \frac{\Gamma(u+1)\Gamma(-u)^{2}}{\Gamma(1-u)} \big(\mu_{t}^{-u} \big(\frac{1}{\epsilon} \big)^{-u+\frac{1}{t}\sum_{i=1}^{N} x_{i}} \Gamma \big(-u+\frac{1}{t}\sum_{i=1}^{N} x_{i} \big) \big) \diff u
\end{eqnarray}
\normalsize
We substitute \eqref{eq:cap_log_int} in \eqref{eq:cap_expn_fso_ris} and apply the definition of $N$-multivariate Fox's H-function \cite[A.1]{M-Foxh} to get \eqref{eq:cap_ris_fso}.
\end{IEEEproof}

We derive an exact closed form expression of the DL-FSO system by substituting \eqref{eq:pdf_combined} in \eqref{eq:gen_cap_fso}, representing $\ln (1+\mu_t\gamma)$ in terms of Meijer's G-function and applying the identity \cite[07.34.21.0012.01]{Meijers}:
		\begin{multline}\label{eq:cap_fso}
		\bar{\eta}^{\rm DL} = \log_{2}(e) \frac{\psi v^k}{t} \sum_{l=1}^{P} \zeta_l C_{l}^{-\phi} \\ H_{p+k+2,q+k+2}^{m+k+2,n+1} \left[C_{l} \bigg(\frac{1}{\mu_t\gamma_0}\bigg)^{1/t}\left|\begin{array}{c}
			V_{1},\big(0,\frac{1}{t}\big),\big(1,\frac{1}{t}\big),V_{2}\\
			V_{3},\big(0,\frac{1}{t}\big),\big(0,\frac{1}{t}\big),V_{4}\\
		\end{array}\right.\right]
	\end{multline}
	\text{where} $V_{1} = \{(\phi+a_{l,w},1)\}_{w=1}^{n}$, $V_{2} = \{(\phi+a_{l,w},1)\}_{w=n+1}^{p},\{(v+1,1)\}_{1}^{k}$, $V_{3} = \{(\phi+b_{l,w},1)\}_{w=1}^{m},\{(v,1)\}_{1}^{k}$ \text{and} $V_{4} = \{(\phi+b_{l,w},1)\}_{w=m+1}^{q}$.
	
\begin{figure*}
	\begin{eqnarray}\label{eq:snr_moments_ris_fso}
		&\bar{\gamma}^{(r)} = \mathbb{E}[\gamma^r] = \frac{1}{t} \bigg(\frac{1}{\epsilon} \bigg)^{r} \sum_{l_{1,1},l_{1,2}=1}^{P}\cdots\sum_{l_{N,1},l_{N,2}=1}^{P}\prod_{i=1}^{N} \prod_{j=1}^{2} \psi_{i,j} v_{i,j}^{k_{i,j}} \zeta_{l_{i,j}} \bigg(C_{l_{i,j}}\bigg)^{-\phi_{i,j}}\nonumber\\& H_{1,1:2p+k_{1,1}+k_{1,2}+1,2q+k_{1,1}+k_{1,2};\cdots;2p+k_{N,1}+k_{N,2}+1,2q+k_{N,1}+k_{N,2}}^{0,1:2m+k_{1,1}+k_{1,2},2n+1;\cdots;2m+k_{N,1}+k_{N,2},2n+1}\left[\begin{array}{c}\bigg(\frac{1}{\epsilon\gamma_0}\bigg)^{1/t} \prod_{j=1}^{2} C_{l_{i,j}}\\.\\.\\.\\\bigg(\frac{1}{\epsilon\gamma_0}\bigg)^{1/t} \prod_{j=1}^{2} C_{l_{i,j}}\end{array} \middle\vert \begin{array}{c}
			(1-r,\frac{1}{t},\cdots,\frac{1}{t}):V_{1} \\
			(1;1,\cdots,1):V_{2}
		\end{array}\right]
	\end{eqnarray}	
	\text{where} $V_{1}=\{\{(\phi_{i,1}+a_{l_{i,1},w},1),\{(\phi_{i,2}+a_{l_{i,2},w},1)\}_{w=1}^{n},(1,1),\{(\phi_{i,1}+a_{l_{i,1},w},1)\}_{w=n+1}^{p},\{(\phi_{i,2}+a_{l_{i,2},w},1)\}_{w=n+1}^{p},\{(v_{i,1}+1,1)\}_{1}^{k_{i,1}},\{(v_{i,2}+1,1)\}_{1}^{k_{i,2}}\}_{i=1}^{N}$ \text{and} $V_{2}=\{\{(\phi_{i,1}+b_{l_{i,1},w},1)\}_{w=1}^{m},\{(\phi_{i,2}+b_{l_{i,2},w},1)\}_{w=1}^{m},\{(v_{i,1},1)\}_{1}^{k_{i,1}},\{(v_{i,2},1)\}_{1}^{k_{i,2}},\{(\phi_{i,1}+b_{l_{i,1},w},1)\}_{w=m+1}^{q},\{(\phi_{i,2}+b_{l_{i,2},w},1)\}_{w=m+1}^{q}\}_{i=1}^{N}$.

	\hrule
\end{figure*}

		\subsection{Moments of SNR}
Finally, we derive moments of SNR for both RISE-FSO and DL-FSO systems, which can be a useful metric  to characterize the average SNR and order of fading.

		\begin{my_lemma} 
	The $r$-th moment of SNR for the RISE-FSO system  is given as  \eqref{eq:snr_moments_ris_fso}.

			\end{my_lemma}
		\begin{IEEEproof} We substitute the PDF of SNR \eqref{eq:pdf_snr_ris} (using \eqref{eq:PDF_final}) in $\mathbb{E}[\gamma^r] = \int_{0}^{\infty} \gamma^r f_{\gamma_{}}(\gamma) \diff \gamma$ to compute the  $r$-th moment of SNR by expanding the definition of Fox's H-function:
\small
	\begin{eqnarray}\label{eq:moment_proof1}
		&\bar{\gamma}^{(r)} = \frac{1}{t} \sum_{l_{1,1},l_{1,2}=1}^{P}\cdots\sum_{l_{N,1},l_{N,2}=1}^{P}\prod_{i=1}^{N} \prod_{j=1}^{2} \psi_{i,j} v_{i,j}^{k_{i,j}} \zeta_{l_{i,j}} \nonumber \\ &\big(C_{l_{i,j}}\big)^{-\phi_{i,j}} \bigg(\frac{1}{2\pi \J}\bigg)^{N} \int\limits_{\mathcal{L}_{i}} \bigg(\bigg(\frac{1}{\gamma_0}\bigg)^{1/t}\prod_{j=1}^{2} C_{l_{i,j}}\bigg)^{x_{i}} \nonumber \\ &\hspace{-6mm}\Bigg[\frac{\prod_{j=1}^{2}\prod_{w=1}^{m} \Gamma\big(-x_{i}+\phi_{i,j}+b_{l_{i,j},w}\big) }{\prod_{j=1}^{2}\prod_{w=n+1}^{p} \Gamma\big(-x_{i}+\phi_{i,j}+a_{l_{i,j},w}\big) } \frac{ \prod_{j=1}^{2}\prod_{w=1}^{n} \Gamma\big(1+x_{i}-\phi_{i,j}-a_{l_{i,j},w}\big)}{\prod_{j=1}^{2}\prod_{w=m+1}^{q} \Gamma\big(1+x_{i}-\phi_{i,j}-b_{l_{i,j},w}\big)} \nonumber \\ & \hspace{-6mm}\frac{\prod_{j=1}^{2}\big(\Gamma\big(v_{i,j}-x_{i}\big)\big)^{k_{i,j}}\Gamma(x_{i})}{\prod_{j=1}^{2}\big(\Gamma\big(1+v_{i,j}-x_{i}\big)\big)^{k_{i,j}}\Gamma\big(\sum_{i=1}^{N}x_i\big)} \Bigg] \bigg(\int_{0}^{\infty} \gamma^{r} \gamma^{-1+\frac{1}{t}\sum_{i=1}^{N} x_{i}}  d\gamma \bigg) \diff x_i			
	\end{eqnarray}
\normalsize
To solve inner integral in \eqref{eq:moment_proof1} we use the  final value theorem:
	\begin{equation}\label{eq:moment_proof2}
		\int_{0}^{\infty} \bigg(\gamma\bigg)^{r-1+\frac{1}{t}\sum_{i=1}^{N} x_{i}} d\gamma = \bigg(\frac{1}{\epsilon} \bigg)^{r+\frac{1}{t}\sum_{i=1}^{N} x_{i}} \Gamma \bigg(r+\frac{1}{t}\sum_{i=1}^{N} x_{i} \bigg)
	\end{equation}
	We substitute \eqref{eq:moment_proof2} in \eqref{eq:moment_proof1} and use the definition of $N$-multivariate Fox's H-function \cite[A.1]{M-Foxh} to get \eqref{eq:snr_moments_ris_fso}.
	\end{IEEEproof}
\normalsize

Similarly, we derive an exact closed form expression of the $r$-th moment of SNR for the DL-FSO system by substituting \eqref{eq:pdf_combined} in  $\mathbb{E}[\gamma^r] = \int_{0}^{\infty} \gamma^r f_{\gamma_{}}(\gamma) \diff \gamma$, expand the Fox's H-function, use the final value theorem to compute the inner integral $\int_{0}^{\infty} \gamma^{r-1+\frac{s}{t}} d\gamma = \lim_{s\rightarrow 0} \big(\frac{1}{s} \big)^{r+\frac{x}{t}} \Gamma \big(r+\frac{x}{t} \big) = \big(\frac{1}{\epsilon} \big)^{r+\frac{x}{t}} \Gamma \big(r+\frac{x}{t} \big)$ to get  \eqref{eq:snr_moments_fso}.
\begin{figure*}
\begin{eqnarray}\label{eq:snr_moments_fso}
	\bar{\gamma}^{(r, \rm DL)} = \mathbb{E}[\gamma^r] = \frac{\psi v^k}{t} \bigg(\frac{1}{\epsilon} \bigg)^{r} \sum_{l=1}^{P} \zeta_l C_{l}^{-\phi} H_{p+k+1,q+k}^{m+k,n+1}\nonumber \\
	\left[C_{l} \bigg(\frac{1}{\epsilon\gamma_0}\bigg)^{1/t}\left|\begin{array}{c}
		(1-r,\frac{1}{t}),\{(\phi+a_{l,w},1)\}_{w=1}^{n},\{(\phi+a_{l,w},1)\}_{w=n+1}^{p},\{(v+1,1)\}_{1}^{k} \\
		\{(\phi+b_{l,w},1)\}_{w=1}^{m},\{(v,1)\}_{1}^{k},\{(\phi+b_{l,w},1)\}_{w=m+1}^{q}\\
	\end{array}\right.\right]	
\end{eqnarray}
\hrule
\end{figure*}

In what follows, we demonstrate the performance of FSO systems using numerical and simulation analysis.

\begin{table*}[h]	
	\caption{Simulation Parameters}
	\label{table:simulation_parameters}
	\centering
	\begin{tabular}{|c|c|c|c|}
		\hline 	
		Transmitted power, $P_T$ & $0$ to $40$ \mbox{dBm} & Responsitivity,	$R$ & $0.41$ \mbox{A/W}\\ \hline
		AWGN variance, $\sigma_{\nu}^2$ & $10^{-14}~\rm {A^2/GHz}$ & Link distance, \{$d$\} & \{$1$ \mbox{km}, 2 \mbox{km}\}\\ \hline	
		Visibility range,  $V$ & $2$ \mbox{km} & Aperture diameter, $D=2a_r$ & $20$ \mbox{cm} \\ \hline	
			Shape parameter of fog, $k$ & \{2, 5\} & 	Scale parameter of fog, $\beta^{\rm fog}$ & \{13.12, 12.06\}\\ \hline	
		Normalized beam-width, $w_z/a_r$ & \{10, 15\} & Normalized jitter, $\sigma_{s}/a_r$ & 3\\ \hline 
		Pointing error angle   $\sigma_{\theta}$ & $1$ \mbox{mrad} & RIS jitter angle  $\sigma_{\beta}$ & $0.5$ \mbox{mrad}\\ standard deviation, & & standard deviation, & \\ \hline
		Modulation,  DBPSK & $p=1, q=1$ & Wavelength, $\lambda$ & $1550$ \mbox{nm}\\ \hline 
		Refractive index, $C_n^2$ & \{$5 \times10^{-14}$,  & $\cal{F}$-distribution, $\{\alpha_{{\scriptscriptstyle F}}, \beta_{{\scriptscriptstyle F}}\}$& \{4.85, 6.55\}, \{17.21, 19.24\}\\ & $1.25 \times10^{-14}$\} \mbox{$m^{-2/3}$} & &\{17.48, 17.8\}, \{68.14, 64.73\}\\ \hline
		$\cal{GG}$-distribution, $\{\alpha_{{\scriptscriptstyle G}}, \beta_{{\scriptscriptstyle G}}\}$& \{3.01, 3\},  \{8.9, 12\}, & $\cal{M}$-distribution, & \{3.01, 3, 0.4, 0.3, 0.596\},  \\& \{8.17, 11\},  \{30.76, 40\} & $\{\alpha_{{\scriptscriptstyle M}}, \beta_{{\scriptscriptstyle M}}, \Omega, b_{0}, \rho\}$& \{8.9, 12, 0.4, 0.3, 0.596\},  \\ & & & \{8.17, 11, 0.4, 0.3, 0.596\}, \\ & & & \{30.76, 40, 0.4, 0.3, 0.596\}\\ \hline		
	\end{tabular}	 
\end{table*} 

\section{Simulation and Numerical Results}\label{sec:sim_num_res}
In this section, we use numerical analysis and Monte Carlo simulations (averaged over $10^{8}$ channel realizations) to demonstrate the performance of the proposed  RISE-FSO system under the combined effect of atmospheric turbulence and pointing errors over different weather conditions. We also compare  the performance of   DL-FSO and RISE-FSO systems using both HD and IM/DD detection techniques. We evaluate the derived analytical expressions using the Python code implementation of multivariate Fox’s H-function \cite{Alhennawi2016} and validate  them through extensive numerical and simulation results. The computational complexity of  Fox's H-function depends more on the number of contour integrals than the computation of integrand involving Gamma functions. Thus, numerical evaluation  of the single-variate Fox's H-function is fast since it requires the computation of  a single contour involving the ratio of products of $m+n$ and $p+q$ Gamma functions even for   large values of $m,n,p,~{\rm{and}}~q$. However, the computation of an $N$-variate Fox's H-function becomes slower with an increase in the number of contour integrals, $N$.    We assume  link distances of $d=1\mbox{km}$ and $d=2\mbox{km}$ for weather conditions of fog and haze, respectively. We assume that the optical RIS is situated  midway between  the source and the destination i.e., $d_{1}=d_{2}=d/2$.  We use parametric equations to  compute $\cal{F}$-turbulence parameters from \cite{Badarneh2020}, $\cal{GG}$ parameters from  \cite{Gappmair2011}, and use $\alpha_{{\scriptscriptstyle M}}=\alpha_{{\scriptscriptstyle G}}$ and $\beta_{{\scriptscriptstyle M}}=\beta_{{\scriptscriptstyle G}}$ for the $\cal{M}$ distribution with $\Omega $, $b_0$, and $\rho$ \cite{Ansari2016}.  We use the recent paper \cite{Wang2020} to model   pointing errors for optical RIS.  We list   simulation parameters  in Table \ref{table:simulation_parameters}.
\begin{figure*}[tp]
	\subfigure[Average SNR.]{\includegraphics[scale=0.5]{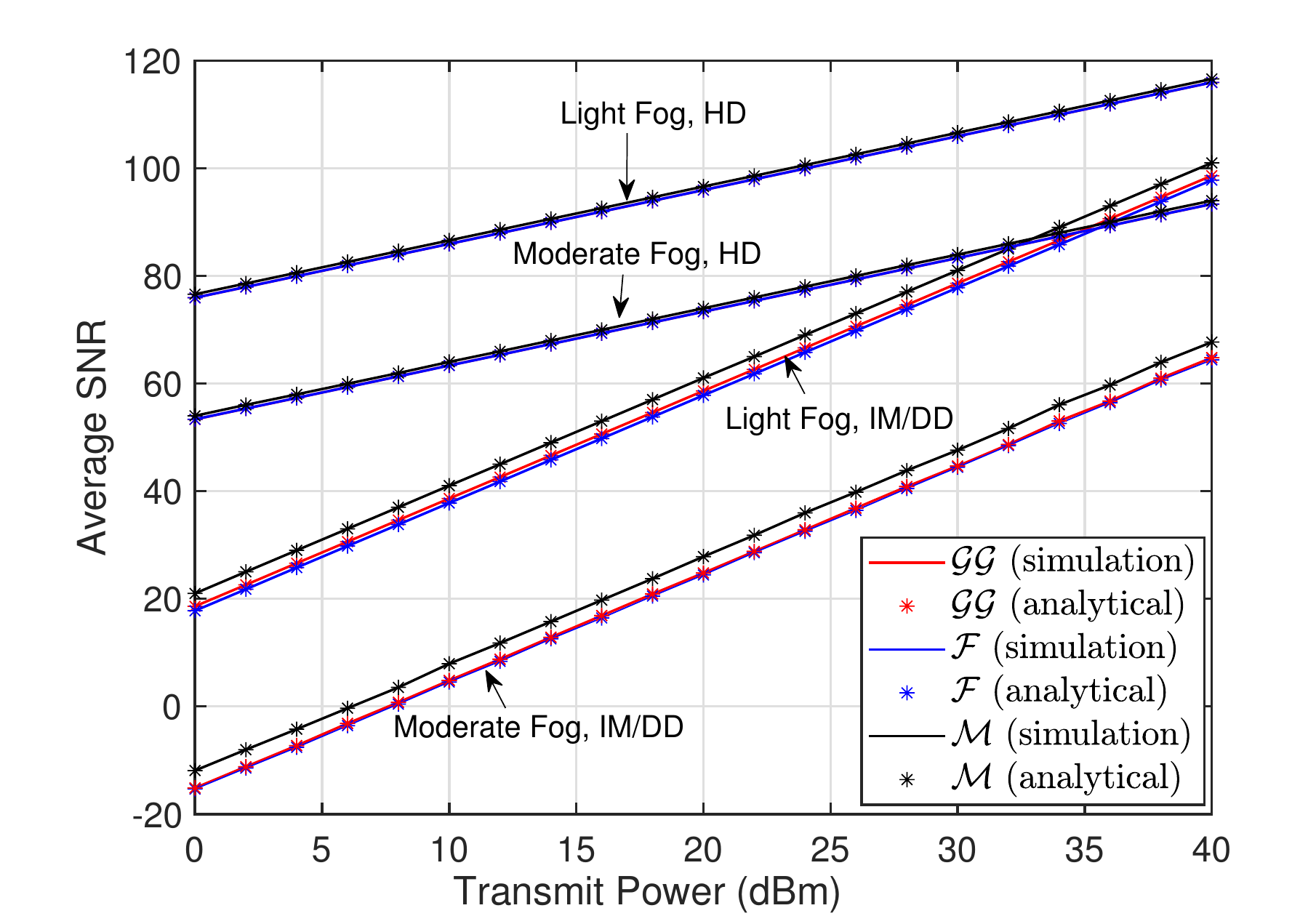}}
	\subfigure[Ergodic capacity.]{\includegraphics[scale=0.5]{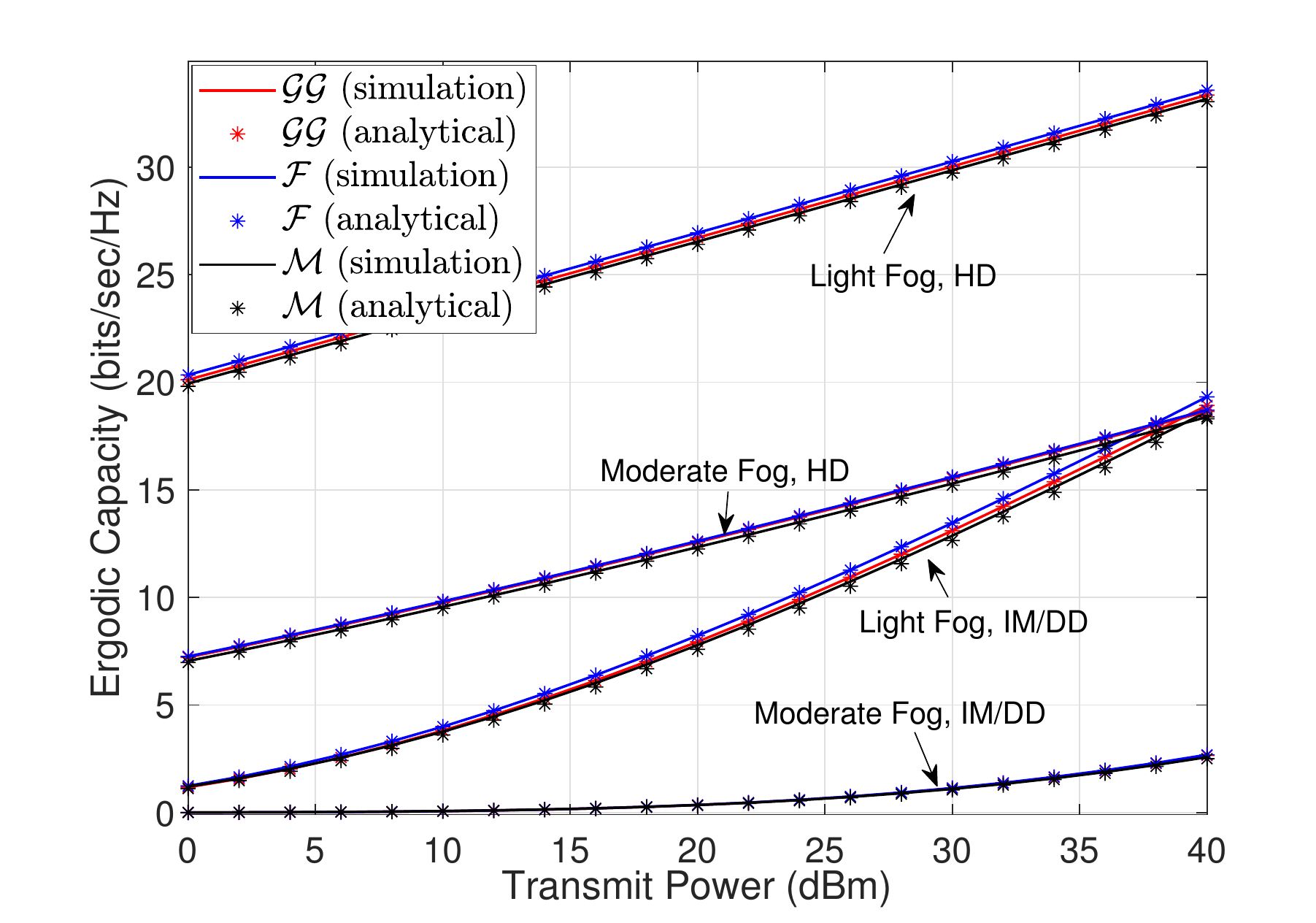}}
	\caption{Average SNR and ergodic capacity of DL-FSO at $d=1\mbox{km}$ for light and moderate fog with HD and IM/DD detection.}
	\label{snr_cap_direct_link}
\end{figure*}
\begin{figure*}[tp]
	\subfigure[Outage probability at  $\gamma_{th}=5\mbox{dB}$.]{\includegraphics[scale=0.5]{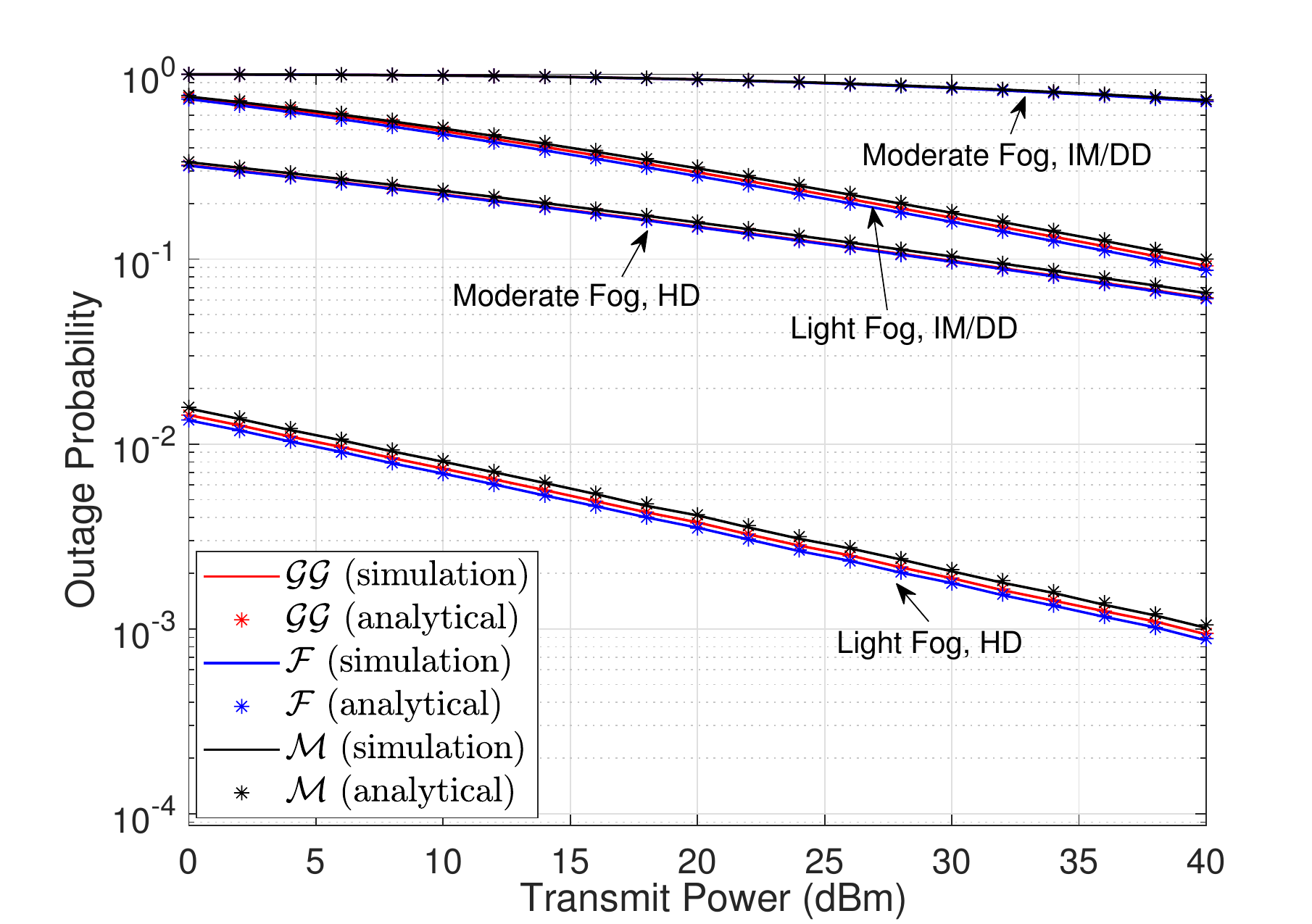}}
	\subfigure[Average BER.]{\includegraphics[scale=0.5]{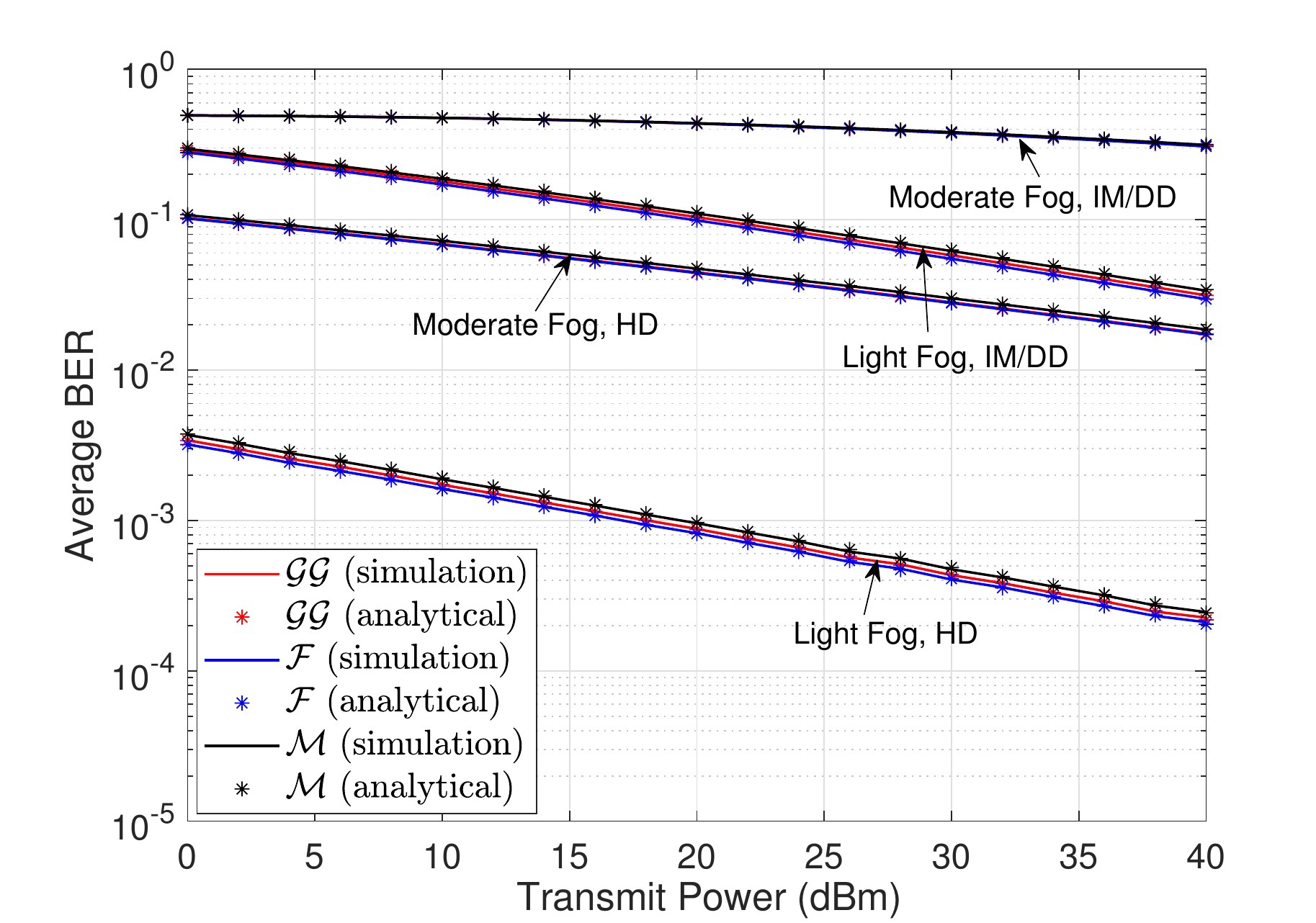}}
	\caption{Outage probability and average BER for DL-FSO system at $d=1\mbox{km}$ with HD and IM/DD detection for light and moderate fog.}
	\label{op_aber_direct_link}
\end{figure*}

 In what follows, we demonstrate the performance of  DL-FSO and RISE-FSO systems in next two subsections.
\subsection{DL-FSO system}
We demonstrate the performance of  DL-FSO system in  Fig.~\ref{snr_cap_direct_link} and Fig.~\ref{op_aber_direct_link} by plotting average SNR, ergodic capacity, outage probability, and average BER with  HD and IM/DD detection under the combined effect of atmospheric turbulence, pointing errors, and foggy conditions. Observing these figures, it can be seen that the  HD detector outperforms IM/DD for the considered atmospheric turbulence models at the expense of decoding complexity.  Fig.~\ref{snr_cap_direct_link}(a) shows that the average SNR is reduced by almost $35$\mbox{dB} and $25$\mbox{dB} at a transmit power of $20$\mbox{dBm} over moderate foggy conditions compared with light foggy for IM/DD and HD detection techniques, respectively. Further, it can be seen from   Fig.~\ref{snr_cap_direct_link}(b) that the moderate fog has a more cumulative impact on IM/DD than the HD  detection. The figure depicts that  the ergodic capacity  has a factor of $6$ reductions for the IM/DD but  with a factor $2$ reduction comparing  light foggy weather to the moderate fog at a transmit power of  $40\mbox{dBm}$. It can also be seen that the performance of FSO system is significantly degraded for back-haul applications over moderate foggy conditions, especially with the IM/DD technique.
\begin{figure*}[tp]
	\subfigure[Average SNR.]{\includegraphics[scale=0.5]{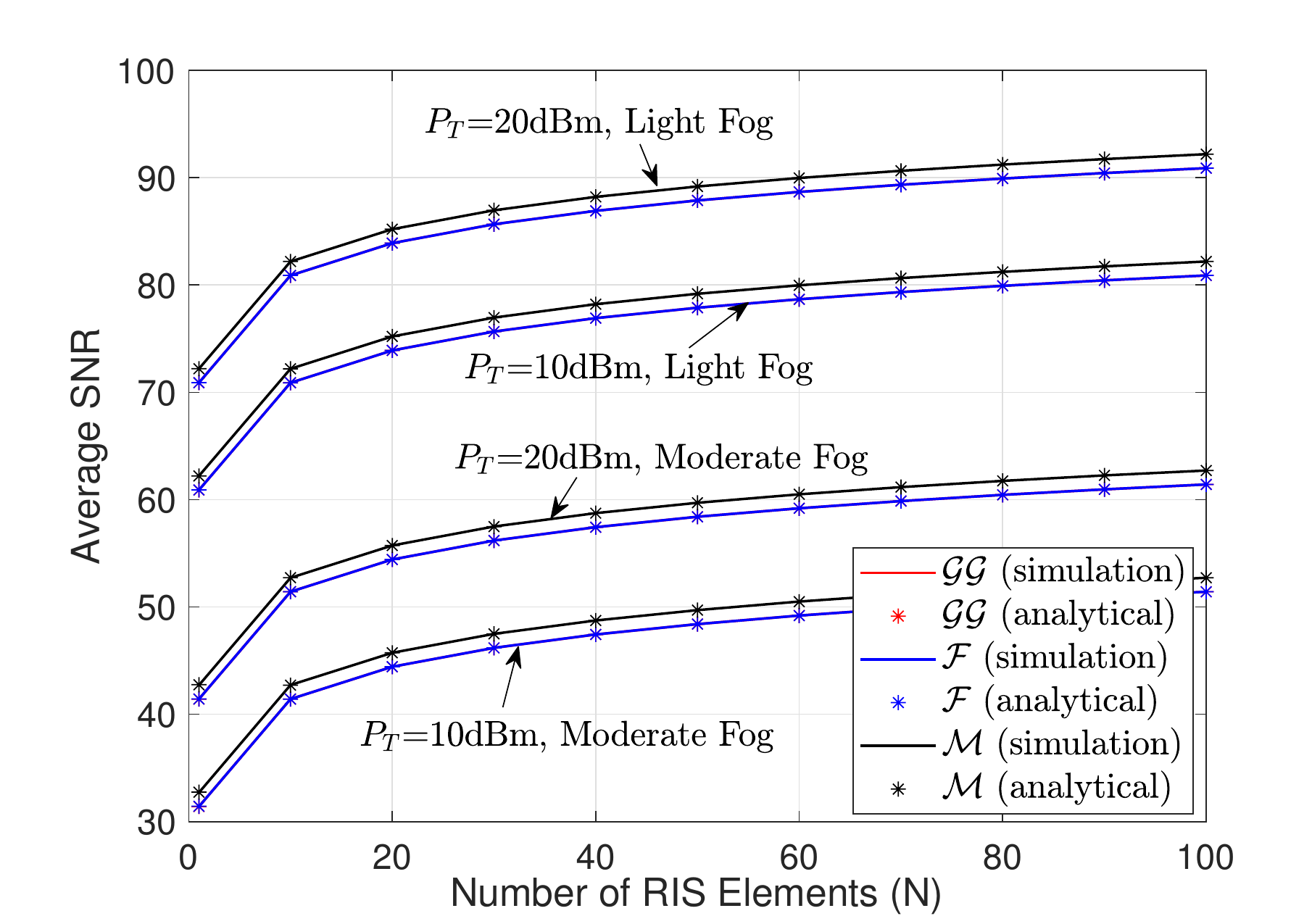}}
	\subfigure[Ergodic capacity.]{\includegraphics[scale=0.5]{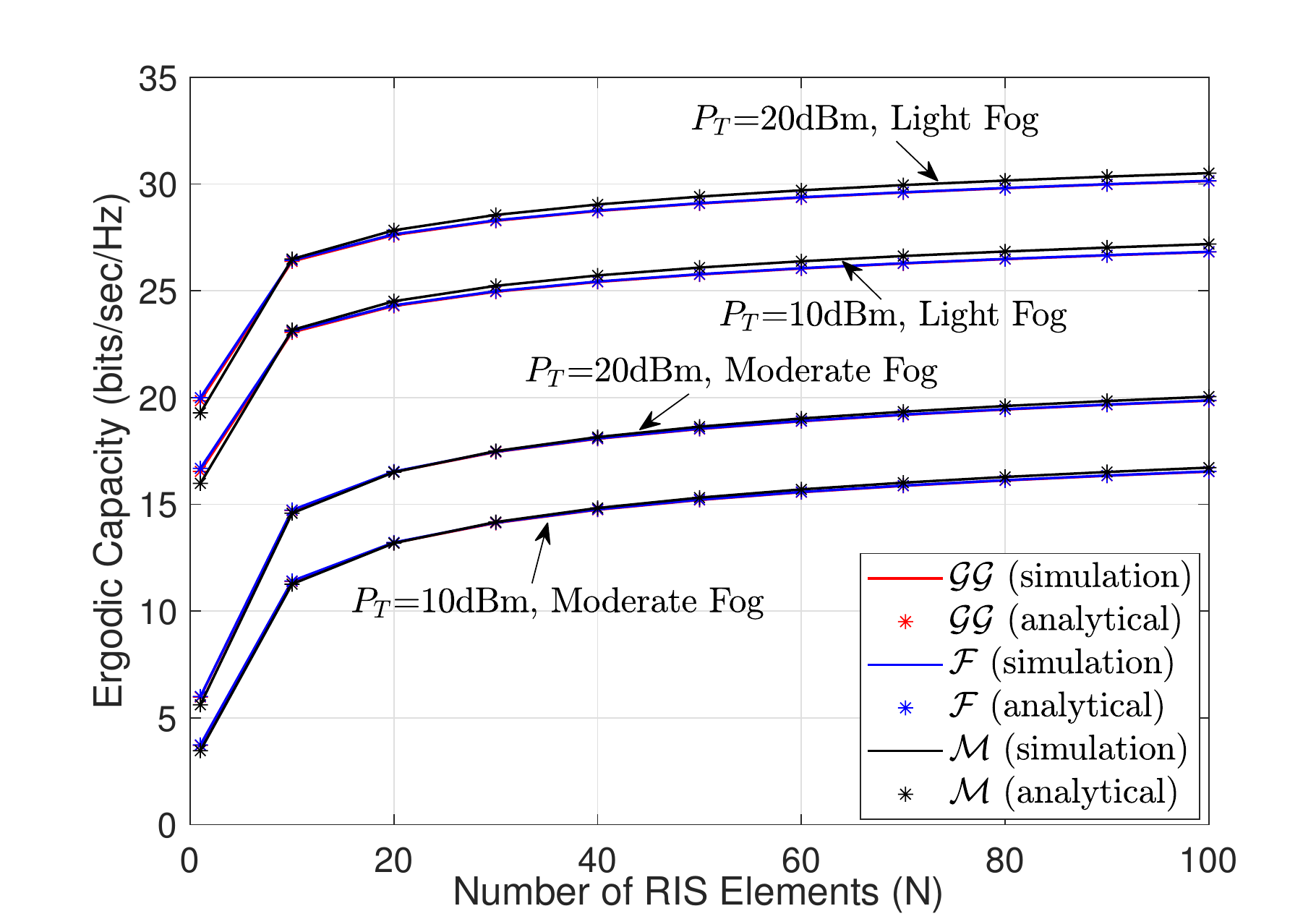}}
	\caption{Average SNR and ergodic capacity of RISE-FSO system at $d_{1}=500\mbox{m}$, and $d_{2}=500\mbox{m}$ with HD for light and moderate fog.}
	\label{snr_cap_ris_fso_fog}
\end{figure*}
\begin{figure*}[tp]
	\subfigure[Outage probability with IM/DD at $\gamma_{th}=5\mbox{dB}$.]{\includegraphics[scale=0.5]{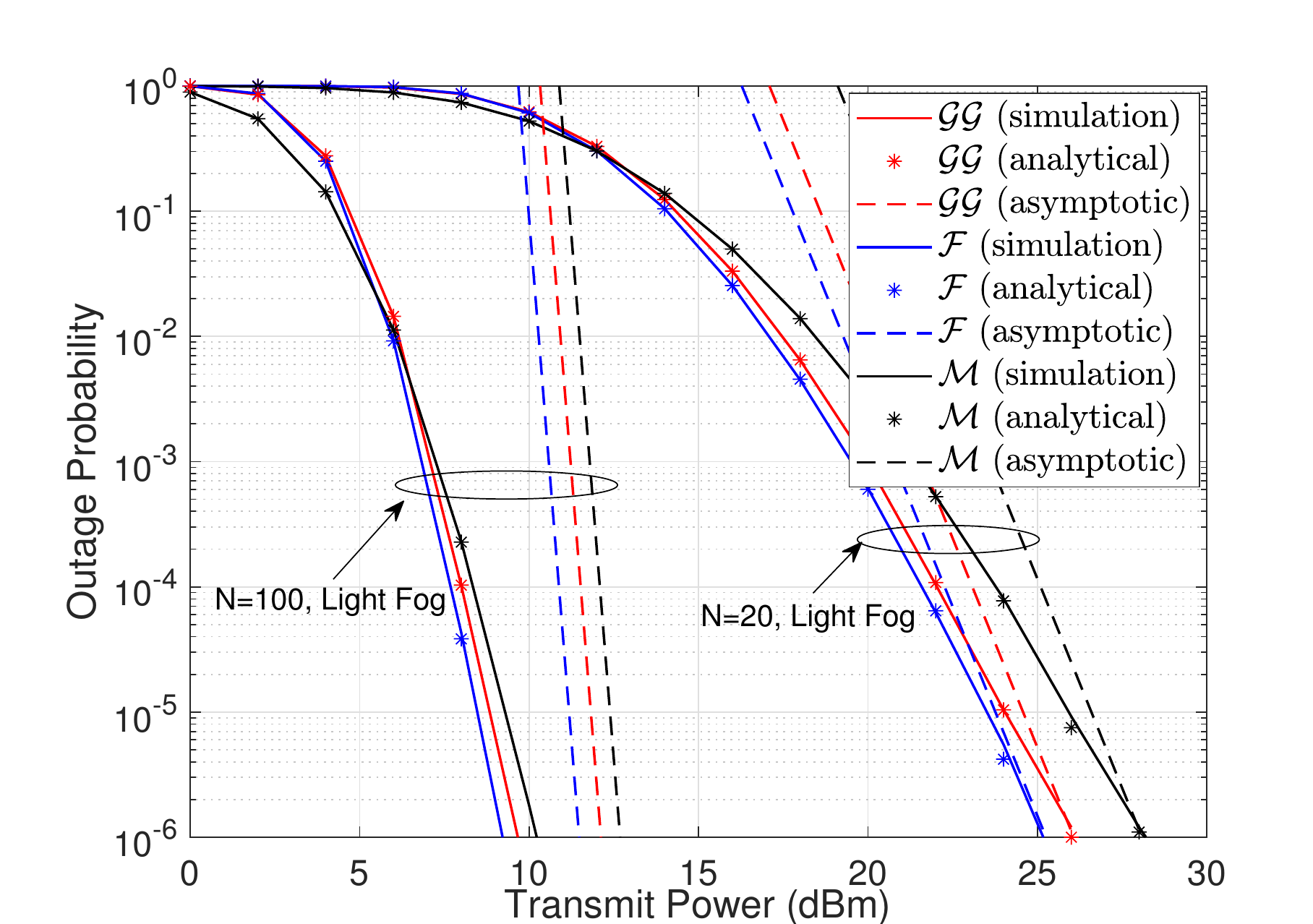}}
	\subfigure[Average BER with IM/DD for light and moderate fog.]{\includegraphics[scale=0.5]{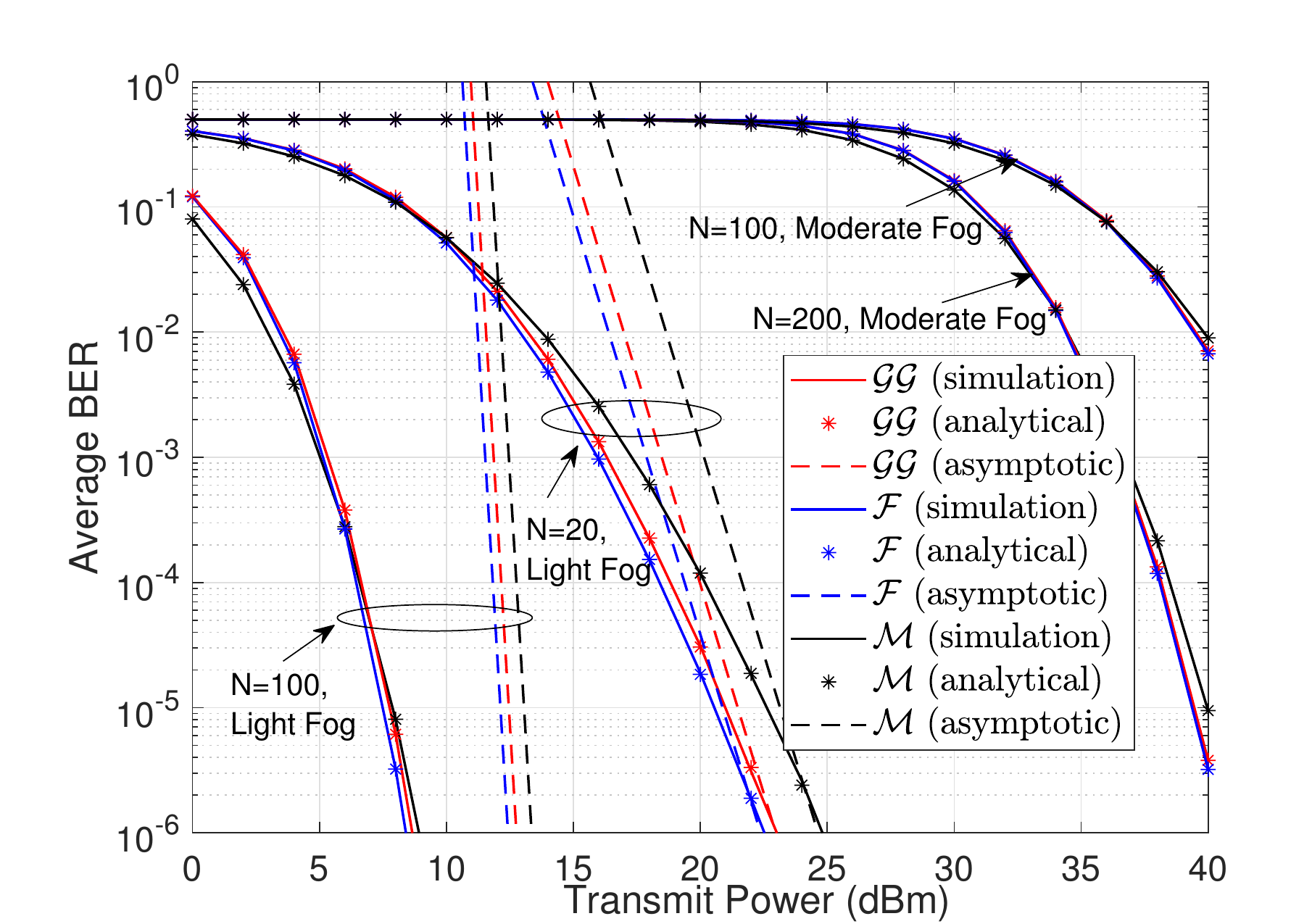}}
	\caption{Outage probability and average BER of RISE-FSO system at $d_{1}=500\mbox{m}$, and $d_{2}=500\mbox{m}$.}
	\label{op_aber_ris_fso_fog}
\end{figure*}
The outage probability and average BER performance of the DL-FSO system  are illustrated in  Fig.~\ref{op_aber_direct_link}. Despite the fact that the HD detector performs better than the IM/DD, the outage probability and average BER performance is significantly degraded to around $10^{-1}$  at a higher transmit $P_{T}=40\mbox{dBm}$ for moderate foggy conditions at a link distance of $1$\mbox{km}. Acceptable reliability of $10^{-3}$ can only be achieved for light foggy conditions with HD detection.  A further decrease in link distance, say $500$\mbox{m},  may improve the reliability of transmissions for a good quality of service.  Considering  parameters of atmospheric turbulence (using Table 	\ref{table:simulation_parameters}) and pointing errors ($\rho^2=2.25$), we can use our analysis to derive the diversity order for the considered DL-FSO system as $\frac{0.33}{t}$ and $\frac{0.36}{t}$ for  light and moderate fog, respectively.    Fig.~\ref{op_aber_direct_link} confirms  the derived diversity order since there is no change in the slope for different turbulence models, almost same slope for similar detection methods for both foggy conditions, and a change  in the slope  comparing the plots for HD ($t=1$)  and IM/DD ($t=2$). Thus, the diversity order provides design criteria to appropriately choose the beamwidth to reduce the impact of pointing errors on FSO systems with other channel impairments.  It can also be seen from Fig.~\ref{op_aber_direct_link}(b) that  analytical expressions and simulation results for the considered binary  modulation scheme have an excellent match over a wide range of SNR. In the following subsection, we employ the optical RIS to improve the performance of the DL-FSO system.

\begin{figure*}[tp]
	\subfigure[Average SNR with HD and IM/DD.]{\includegraphics[scale=0.5]{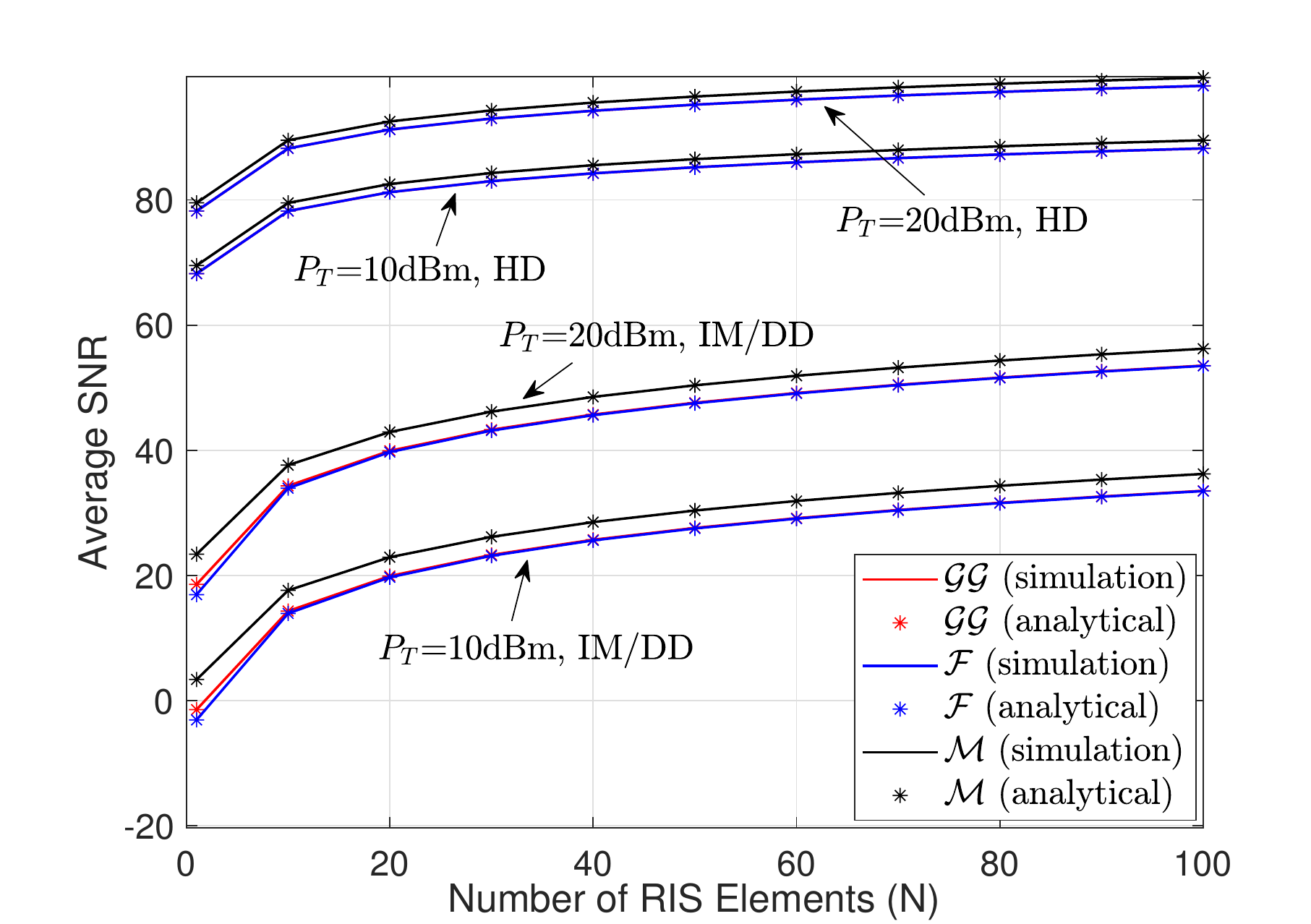}}
	\subfigure[Ergodic capacity with HD and IM/DD.]{\includegraphics[scale=0.5]{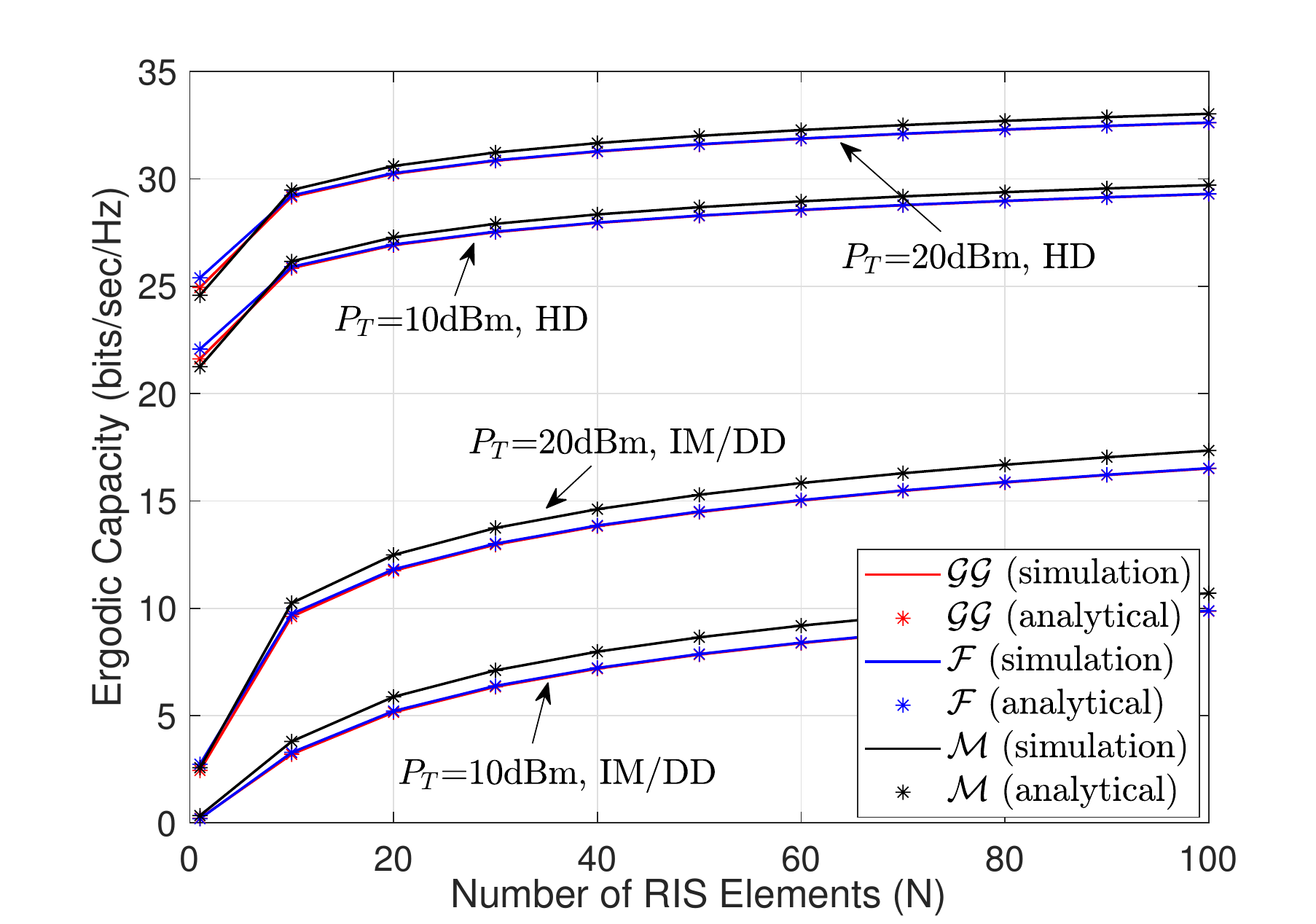}}
	\caption{Average SNR and ergodic capacity of RISE-FSO system with deterministic path loss at $d_{1}=1\mbox{km}$, $d_{2}=1\mbox{km}$, and $C_{n}^{2}=5 \times 10^{-14}$ \mbox{$m^{-2/3}$}.}
	\label{snr_cap_ris_fso}
\end{figure*}

\begin{figure*}[tp]
	\subfigure[Outage probability with  IM/DD at $\gamma_{th}=5\mbox{dB}$.]{\includegraphics[scale=0.5]{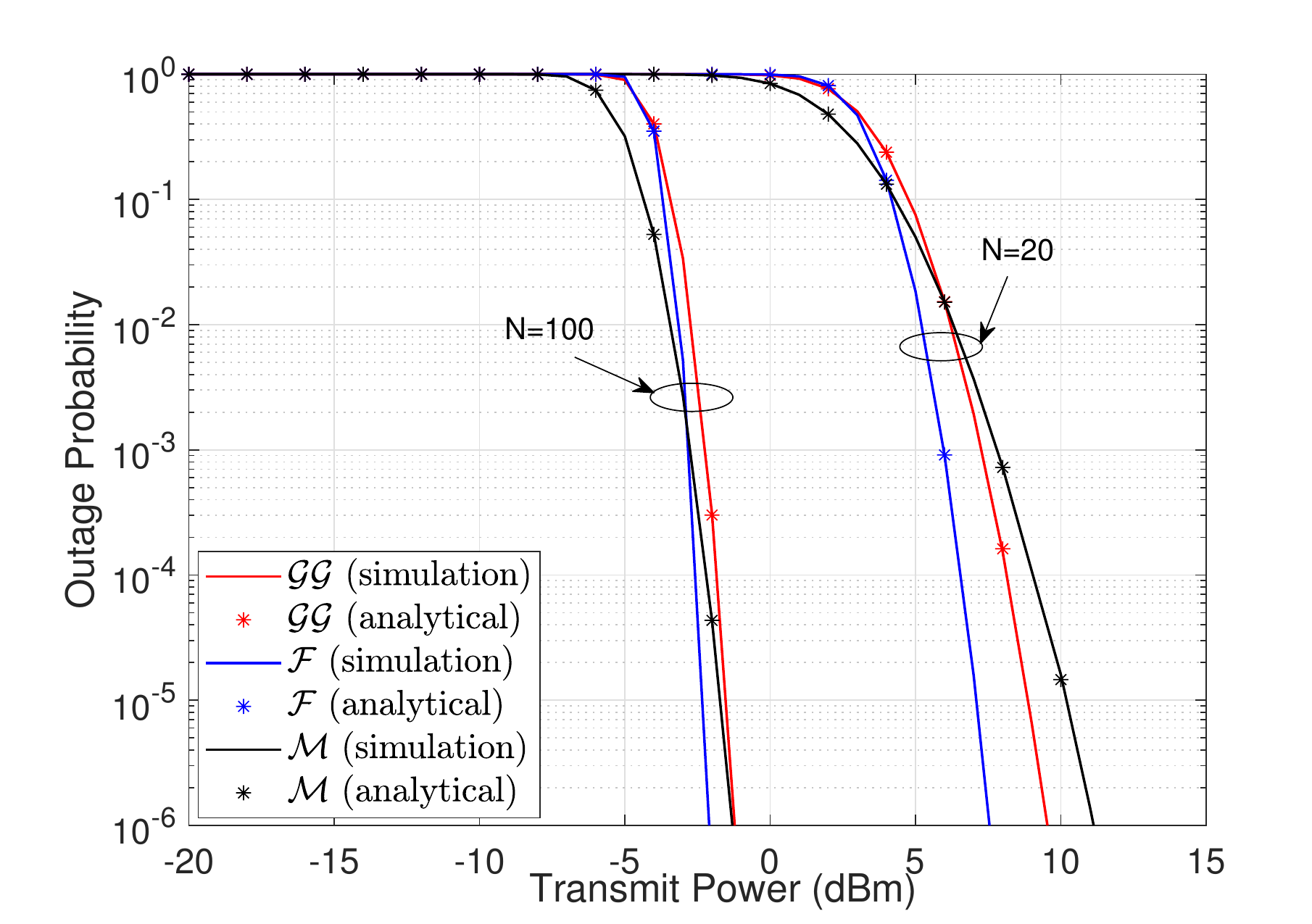}}
	\subfigure[Average BER with IM/DD.]{\includegraphics[scale=0.5]{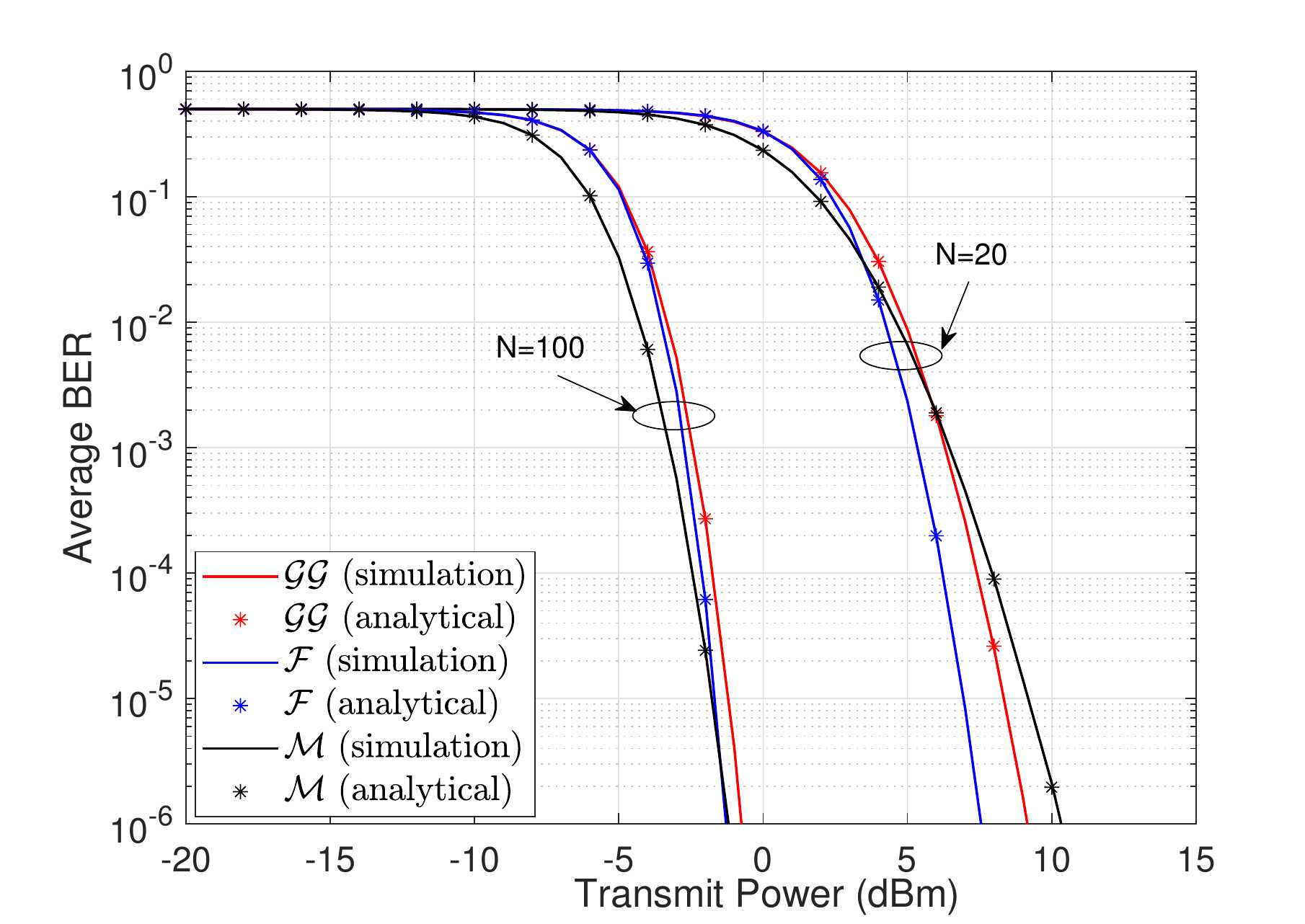}}
	\caption{Outage probability and average BER of RISE-FSO system with deterministic path loss at $d_{1}=1\mbox{km}$, $d_{2}=1\mbox{km}$, and $C_{n}^{2}=5\times10^{-14}$ \mbox{$m^{-2/3}$}.}
	\label{op_aber_ris_fso}
\end{figure*}

\subsection{RISE-FSO system}
We  use parameters for atmospheric turbulence and pointing errors customized for the optical RIS to simulate the  RISE-FSO system, as listed in Table \ref{table:simulation_parameters}. Without loss of generality,  we assume i.i.d  channel model for both the hops by considering the same parameters of the  atmospheric turbulence, random fog, and pointing errors from source to the RIS and RIS to the destination. In Fig.~\ref{snr_cap_ris_fso_fog}, we demonstrate the impact of RIS elements  on the average SNR  and ergodic capacity for light and moderate  fog with HD technique at a transmit power of  $10$\mbox{dBm} and $20$\mbox{dBm}. It can be seen that average SNR and ergodic capacity increase with an increase in the number of RIS elements.  Compared with the RISE-FSO,  the average SNR for the DL-FSO system is higher than the $N=100$ RIS system.  However, comparing the ergodic capacity of RISE-FSO in Fig.~\ref{snr_cap_ris_fso_fog}(b) with the DL-FSO in Fig.~\ref{snr_cap_direct_link}(b), we can observe that a $100$ element RIS surface can provide a significant  increase of $3$ \mbox{bits/sec/Hz} in spectral efficiency.

There is a significantly higher  improvement in the performance of outage probability and average BER with RIS, as shown in Fig.~\ref{op_aber_ris_fso_fog}.  We demonstrate the impact of RIS elements on these performance metrics   for light and moderate  fog conditions with the IM/DD detector.  Fig.~\ref{op_aber_ris_fso_fog}(a) shows that the $N=100$ RISE-FSO achieves a gain of about  $12\mbox{dBm}$  of transmit power to achieve the same outage probability of $10^{-3}$ compared with the $N=20$ RIS system under  light fog conditions. Further, we can verify the impact of RIS elements on the diversity order of RISE-FSO system as predicted through analytical results. In Fig.~\ref{op_aber_ris_fso_fog}(b), we plot the average BER of the RISE-FSO system  for both light and moderate foggy conditions. The figure shows that the average BER improves significantly  with an  increase in the number of RIS elements. Moreover,  the behavior of slope in the average BER with $N$ depicts the diversity gain of the RISE-FSO system.  We also verify the slope of the BER and the outage probability at high SNR by plotting the numerically evaluated derived asymptotic expressions for both outage probability and average BER for light foggy conditions. It can also be seen from Fig.~\ref{op_aber_ris_fso_fog}(b)  that the average BER for the moderate fog condition is still higher even with $N=100$ RIS elements, the performance degradation caused by moderate fog can be compensated with a further increase in $N$ (as seen with the plot $N=200$ in Fig.~\ref{op_aber_ris_fso_fog}(b)). Thus, it is possible to  reduce the effect of atmospheric turbulence, pointing errors,  and adverse weather  conditions by  increasing the number of RIS elements demonstrating a potential design criteria for terrestrial FSO systems.

Finally, we consider the  conventional FSO communications over atmospheric turbulence and pointing errors with deterministic  path loss evaluated using  the well known 
Beer-Lambert's law and visibility range. In Fig.~\ref{snr_cap_ris_fso} and Fig.~\ref{op_aber_ris_fso}, we use similar simulation parameters of the  RISE-FSO with random fog to perform experiments on the  average SNR, ergodic capacity,  and outage probability, average BER of the RISE-FSO  by considering non-foggy condition with a visibility range of $2$ \mbox{km} (a typical haze weather condition).  Fig.~\ref{snr_cap_ris_fso} and Fig.~\ref{op_aber_ris_fso} demonstrate that there is an improvement in the FSO system due to lower path loss with a higher visibility range of haze conditions. Further, figures show that the impact of RIS elements on the performance of RISE-FSO without fog follows a similar trend to that of the  RISE-FSO with fog, as described in the preceding paragraphs.

In all the above figures  (Fig. 2 to Fig. 7), it can be seen that our derived unified expressions have a good agreement with Monte-Carlo simulations validating the proposed analysis. Moreover, the performance of the FSO system is similar for ${\cal{F}}$, ${\cal{GG}}$, and $\cal{M}$ models  since the  turbulence scenarios (i.e., from medium to strong as depicted by the respective turbulence parameters) are applicable for  the three atmospheric models. However, ${\cal{F}}$, ${\cal{GG}}$ slightly overestimates the performance for very strong turbulence  compared with $\cal{M}$-distribution.

\section{Conclusions}
In this paper, we presented exact  closed-form expressions on the performance of RIS empowered FSO system under various channel impairments such as atmospheric turbulence, pointing errors, and different weather conditions.  Our derived analytical results are unified, allowing evaluation of the RISE-FSO system over  $\cal{F}$, $\cal{GG}$, and $\cal{M}$ atmospheric turbulence models with pointing errors,  deterministic and  random path-loss, and considering both HD and IM/DD detection techniques. We developed an exact analysis of the performance metrics such as outage probability, average BER, ergodic capacity, and moments of SNR of the RISE-FSO system.  Using the  asymptotic analysis on the outage probability and  average BER,  we derived the diversity order , which provides different design criteria to circumvent the effect of pointing errors and  random fog for the FSO system under atmospheric turbulence using the proposed RIS based solution.  As such, an increase in the RIS elements significantly  improve the FSO performance with non-LOS transmission link, whereas the use of  suitable beamwidth mitigates the impact of pointing errors, and limiting the communication range is useful for reducing the effect of random fog.  We provided extensive simulations and numerical analysis to demonstrate the effectiveness of the RISE-FSO system comparing with the DL-FSO under various channel conditions. Simulation plots provide various design configurations of  system and channel parameters to achieve the desired performance. It has been shown that the performance degradation caused by atmospheric turbulence, pointing errors,  and adverse weather  conditions  can be compensated by  increasing the number of RIS elements. The proposed work demonstrated the application of optical RIS  to enhance the performance of the FSO system considering a  general scenario of atmospheric  turbulence,  pointing error impairments, weather conditions, and the underlying detection methods. 
We envision that the  RIS technology can empower the deployment of the FSO system for next generation wireless networks, especially for terrestrial  applications.

As a future scope, it would be interesting to combine the RISE-FSO system with RIS-assisted RF for  better connectivity in the access network. Further, we may extend the single-RIS based system  to multi-RIS for  enhanced performance. Analysis of  the system performance with imperfect phase compensation at the RIS may also be conducted.

\section*{Appendix A: PDF and CDF of Direct Link $h_i$}
	 We use the joint distribution of conditional random variables to get the PDF of  $h_i=h_i^{(tp)}h_i^{(f)}$  as  \cite{Papoulis2001}:
\begin{equation}
	\label{eq:pdf_cond}
	f_{h_{i}}(x)=\int_{x}^{\infty} \frac{f_{h_{i}^{(f)}}(\frac{x}{u})f_{h_{i}^{(tp)}}(u)}{u}\diff u
\end{equation}
where the limits of the integral are selected  using the inequalities 	$0\leq \frac{x}{u}\leq 1$ and $0\leq u\leq \infty$ since  $h_{i}^{(f)}\in [0,1]$ and $h_{i}^{(t)}\in [0,\infty)$. Using \eqref{eq:h_fg} and \eqref{eq:PDF_gen_pe} in 	\eqref{eq:pdf_cond}, we get
\begin{eqnarray}
	\label{eq:pdf_cond2}
	f_{h_{i}}(x)=\psi \frac{v^k}{\Gamma(k)} \sum_{l=1}^{P} \zeta_{l} x^{v-1}\int_x^\infty u^{\phi-1}\frac{\ln^{k-1}(\frac{u}{x})}{u^v}\nonumber\\
	G_{p,q}^{m,n}\left[C_{l} u \left|
	\begin{array}{c}
		\{a_{l,w}\}_{w=1}^{p} \\ 
		\{b_{l,w}\}_{w=1}^{q}\\
	\end{array}
	\right.\right]\diff u
\end{eqnarray}

We use the definition of Meijer's G-function, interchange the order of integration to express	\eqref{eq:pdf_cond2} as
\begin{eqnarray}
	\label{eq:pdf_cond3}
	f_{h_{i}}(x) = \psi \frac{v^k}{\Gamma(k)} \sum_{l=1}^{P} \zeta_{l} x^{v-1}\frac{1}{2\pi \J}\int\limits_{\mathcal{L}} \bigg(C_{l} \bigg)^{s} \frac{\prod_{j=1}^{m}\Gamma(b_{l,j}-s)}{\prod_{j=n+1}^{p}\Gamma(a_{j}-s)} \nonumber \\ \frac{\prod_{j=1}^{n}\Gamma(1-a_{l,j}+s)}{\prod_{j=m+1}^{q}\Gamma(1-b_{l,j}+s)}  \bigg(\int_x^\infty u^{\phi-v-1}\ln^{k-1}\bigg(\frac{u}{x}\bigg) u^{s} \diff u\bigg)\diff s
\end{eqnarray}
Substituting $\ln(\frac{u}{x})=y$ and applying 	$\frac{1}{v-s-\phi} = \frac{\Gamma(v-s-\phi)}{\Gamma(v-s-\phi+1)}$,  the inner integral in \eqref{eq:pdf_cond3} can be solved as
\begin{eqnarray}
	\label{eq:pdf_cond4}
	\hspace{-6mm}\int_x^\infty u^{s+\phi-v-1}\ln^{k-1}\bigg(\frac{u}{x}\bigg) \diff u \nonumber \\ = \frac{x^{s+\phi-v}(\Gamma(v-s-\phi+1))^k}{(\Gamma(v-s-\phi))^{k}} \Gamma(k)
\end{eqnarray}

Finally, we use 	\eqref{eq:pdf_cond4} in 		\eqref{eq:pdf_cond3}	and apply the definition of Meijer's G-function, we get (\ref{eq:pdf_combined}) of Theorem 1. To derive the CDF, we use the following:
\small
\begin{eqnarray}
	\label{eq:cond_pdf5}
	F_{h_i}(x) = \int_{0}^{x} f_{h_i}(u) \diff u = \psi v^k \sum_{l=1}^{P} \zeta_l \nonumber \\ \frac{1}{2\pi \J} \int\limits_{\mathcal{L}} \frac{\prod_{j=1}^{m}\Gamma(b_{l,j}-s)\prod_{j=1}^{n}\Gamma(1-a_{l,j}+s)}{\prod_{j=n+1}^{p}\Gamma(a_{l,j}-s)\prod_{j=m+1}^{q}\Gamma(1-b_{l,j}+s)} \nonumber \\ \frac{\big(\Gamma(v- s-\phi)\big)^{k}}{\big(\Gamma(v-s-\phi+1)\big)^{k}} \big(C_l \big)^{s} \big(\int_{0}^{x} v^{ s+\phi-1} \diff v\big)\diff s 
\end{eqnarray}
\normalsize
Using the solution of  inner integral $\int_{0}^{x} v^{ s+\phi-1} \diff v = \frac{x^{s+\phi}}{s+\phi}=x^{s+\phi}\frac{\Gamma(s+\phi)}{\Gamma(s+\phi+1)}$ in 	\eqref{eq:cond_pdf5}, we  apply the definition of Meijer's G-function to get (\ref{eq:cdf_combined}).

	As a sanity check, we verify the derived PDF by considering the $\cal{F}$-turbulence scenario. Thus, we use parameters from Table \ref{table:unified_param} for the $\cal{F}$-distribution and  apply the  identity  \cite[07.34.21.0009.01]{Meijers}:
\small
				\begin{align}	
			&	\int_{0}^{\infty} f_{h_{i}}(x) \diff x = \frac{\alpha_{{\scriptscriptstyle F}}\rho^2v^k}{(\beta_{{\scriptscriptstyle F}}-1)A_0\Gamma(\alpha_{{\scriptscriptstyle F}})\Gamma(\beta_{{\scriptscriptstyle F}})} \nonumber \\ &\int_{0}^{\infty} G_{k+2,k+2}^{k+2,1} \left[\frac{\alpha_{{\scriptscriptstyle F}}}{(\beta_{{\scriptscriptstyle F}}-1)A_0}x\left|\begin{array}{c}
				-\beta_{{\scriptscriptstyle F}},\rho^2,\{v\}_{1}^{k} \\
				\alpha_{{\scriptscriptstyle F}}-1,\rho^2-1,\{v-1\}_{1}^{k}\\
			\end{array}\right.\right] \diff x \nonumber \\
			&= \frac{\alpha_{{\scriptscriptstyle F}}\rho^2v^k}{(\beta_{{\scriptscriptstyle F}}-1)A_0\Gamma(\alpha_{{\scriptscriptstyle F}})\Gamma(\beta_{{\scriptscriptstyle F}})} \frac{\Gamma(\alpha_{{\scriptscriptstyle F}})\Gamma(\rho^{2})\big(\Gamma(v)\big)^{k}\Gamma(\beta_{{\scriptscriptstyle F}})}{\Gamma(1+\rho^{2})\big(\Gamma(1+v)\big)^{k}} \frac{(\beta_{{\scriptscriptstyle F}}-1)A_0}{\alpha_{{\scriptscriptstyle F}}} \nonumber \\ &= 1
		\end{align}
\normalsize

  \section*{Appendix B: PDF, CDF, and MGF of $Z_i$}
We  use the Mellin transform to derive the PDF of $Z_{i} = \prod_{j=1}^{L}h_{i, j}$. Here, $h_{i, j}, j= 1, 2,\cdots,L$ are considered to be i.ni.d random variables  distributed according to (\ref{eq:pdf_combined}). Thus, the PDF of $Z_i$:
\begin{equation}\label{eq:Mellin_pdf}
	f_{Z_{i}}(x) = \frac{1}{x} \frac{1}{2\pi \J} \int\limits_{\mathcal{L}} \mathbb{E}[Z_{i}^r] x^{-r} \diff r
\end{equation}
where  $\mathbb{E}[Z_{i}^r] = \prod_{j=1}^{L} \mathbb{E}[h_{i,j}^{r}] = \prod_{j=1}^{L} \int_{0}^{\infty} u^r f_{h_{i,j}}(u) \diff u$ is the $r$-th moment of $Z_i$.  
Here,  ${\cal{L}}$ is an infinite contour in the complex  $r$-plane such that the integrand in \eqref{eq:Mellin_pdf} has no singularities \cite{M-Foxh}.
 We substitute the PDF of $h_{i,j}$ and use the identity \cite[07.34.21.0009.01]{Meijers} to get  
\small
	\begin{align}
		&\int_{0}^{\infty} u^r f_{h_{i,j}}(u) \diff u = \psi_{j} v_{j}^{k_{j}} \sum_{l_{j}=1}^{P} \zeta_{l_{j}}   \int_{0}^{\infty} u^r   u^{\phi_{j}-1}  \nonumber \\ &G_{p+k_{j},q+k_{j}}^{m+k_{j},n}\left[C_{l_{j}} u\left|\begin{array}{c}
			\{a_{l+{j},w}\}_{w=1}^{p},\{v-\phi_{j}+1\}_{1}^{k_{j}} \\
			\{b_{l_{j},w}\}_{w=1}^{m},\{v-\phi_{j}\}_{1}^{k_{j}},\{b_{l_{j},w}\}_{w=m+1}^{q}\\
		\end{array}\right.\right]\diff u  \nonumber \\
	  &= \psi_{j} v_{j}^{k_{j}} \sum_{l_{j}=1}^{P} \zeta_{l_{j}} \big(C_{l_{j}}\big)^{-(r+\phi_{j})} \Bigg[\frac{\prod_{w=1}^{m} \Gamma\big(r+\phi_{j}+b_{l_{j},w}\big) }{\prod_{w=r+1}^{p} \Gamma\big(r+\phi_{j}+a_{l_{j},w}\big) } \nonumber \\ & \frac{\prod_{w=1}^{n} \Gamma\big(1-r-\phi_{j}-a_{l_{j},w}\big)\big(\Gamma\big(v_{j}+r\big)\big)^{k_{j}}}{ \prod_{w=m+1}^{q} \Gamma\big(1-r-\phi_{j}-b_{l_{j},w}\big)\big(\Gamma\big(1+v_{j}+r\big)\big)^{k_{j}}}\Bigg] 
	\end{align}
\normalsize
Thus, the	$r$-th moment of $Z_{i}$ is given by
\small
	\begin{flalign}\label{eq:moment_appendix}
		&\mathbb{E}[Z_{i}^r] = \sum_{l_{1},\cdots,l_{L}=1}^{P} \prod_{j=1}^{L} \psi_{j} v_{j}^{k_{j}} \zeta_{l_{j}} \prod_{j=1}^{L}\bigg(C_{l_{j}}\bigg)^{-(r+\phi_{j})} \nonumber \\
		&\Bigg[\frac{\prod_{j=1}^{L}\prod_{w=1}^{m} \Gamma\big(r+\phi_{j}+b_{l_{j},w}\big) }{\prod_{j=1}^{L}\prod_{w=n+1}^{p} \Gamma\big(r+\phi_{j}+a_{l_{j},w}\big) } \nonumber \\ &\frac{ \prod_{j=1}^{L}\prod_{w=1}^{n} \Gamma\big(1-r-\phi_{j}-a_{l_{j},w}\big)}{ \prod_{j=1}^{L}\prod_{w=m+1}^{q} \Gamma\big(1-r-\phi_{j}-b_{l_{j},w}\big)} \frac{\prod_{j=1}^{L}\big(\Gamma\big(v_{j}+r\big)\big)^{k_{j}}}{\prod_{j=1}^{L}\big(\Gamma\big(1+v_{j}+r\big)\big)^{k_{j}}}\Bigg]	
	\end{flalign}	
\normalsize	
We substitute \eqref{eq:moment_appendix} in \eqref{eq:Mellin_pdf} to get the PDF of $Z_{i}$ as		
\small
	\begin{align}\label{eq:pdf_prod_ris_appendix}
		&f_{Z_{i}}(x) =\frac{1}{x} \frac{1}{2\pi \J} \sum_{l_{1},\cdots,l_{L}=1}^{P} \prod_{j=1}^{L} \psi_{j} v_{j}^{k_{j}} \zeta_{l_{j}} \bigg(C_{l_{j}}\bigg)^{-\phi_{j}} \int\limits_{\mathcal{L}}  \bigg(x\prod_{j=1}^{L}C_{l_{j}}\bigg)^{-r} \nonumber \\ &\Bigg[\frac{\prod_{j=1}^{L}\prod_{w=1}^{m} \Gamma\big(r+\phi_{j}+b_{l_{j},w}\big) }{\prod_{j=1}^{L}\prod_{w=n+1}^{p} \Gamma\big(r+\phi_{j}+a_{l_{j},w}\big) } \nonumber \\ &\frac{\prod_{j=1}^{L}\prod_{w=1}^{r} \Gamma\big(1-r-\phi_{j}-a_{l_{j},w}\big)\prod_{j=1}^{L}\big(\Gamma\big(v_{j}+r\big)\big)^{k_{j}}}{\prod_{j=1}^{L}\prod_{w=m+1}^{q} \Gamma\big(1-r-\phi_{j}-b_{l_{j},w}\big)\prod_{j=1}^{L}\big(\Gamma\big(1+v_{j}+r\big)\big)^{k_{j}}}\Bigg] \diff r
	\end{align}
\normalsize
The region of convergence of the contour integral ${\mathcal{L}}$ depends on $\arg(x\prod_{j=1}^{L}C_{l_{j}})$ and $ \delta = m+n-\frac{p+q}{2}$, which is the entire plane if $\arg(x\prod_{j=1}^{L}C_{l_{j}})\leq \delta \pi$  \cite[07.34.02.0001.01]{Meijers}. Since $\delta= 1$ and $\arg(x\prod_{j=1}^{L}C_{l_{j}})=0$,   the region of the contour integral in \eqref{eq:pdf_prod_ris_appendix} is $ {\mathcal{L}}:-\J \infty \to +\J \infty $.
	
Hence, we  apply the definition of Meijer's G-function in \eqref{eq:pdf_prod_ris_appendix} to get \eqref{eq:pdf_prod_ris}.	The CDF of $Z_{i}$ can be obtained as $F_{Z_{i}}(x) = \int_{0}^{x} f_{Z_{i}}(u)  \diff u$. Thus, 
\small
	\begin{align}
		\label{eq:appendix_zaf}
		&F_{Z_{i}}(x) = \frac{1}{2\pi \J} \sum_{l_{1},\cdots,l_{L}=1}^{P} \prod_{j=1}^{L} \psi_{j} v_{j}^{k_{j}} \zeta_{l_{j}} \bigg(C_{l_{j}}\bigg)^{-\phi_{j}}\int\limits_{\mathcal{L}}  \bigg(\prod_{j=1}^{L}C_{l_{j}}\bigg)^{r} \nonumber \\ & \bigg(\int_{0}^{x} u^{r-1} \diff u \bigg) \Bigg[\frac{\prod_{j=1}^{L}\prod_{w=1}^{m} \Gamma\big(-r+\phi_{j}+b_{l_{j},w}\big) }{\prod_{j=1}^{L}\prod_{w=n+1}^{p} \Gamma\big(-r+\phi_{j}+a_{l_{j},w}\big) } \nonumber \\ &\frac{\prod_{j=1}^{L}\prod_{w=1}^{n} \Gamma\big(1+r-\phi_{j}-a_{l_{j},w}\big)\prod_{j=1}^{L}\big(\Gamma\big(v_{j}-r\big)\big)^{k_{j}}}{\prod_{j=1}^{L}\prod_{w=m+1}^{q} \Gamma\big(1+r-\phi_{j}-b_{l_{j},w}\big)\prod_{j=1}^{L}\big(\Gamma\big(1+v_{j}-r\big)\big)^{k_{j}}}\Bigg]\diff r		
	\end{align}
\normalsize
	Using the  inner integral solved by the identity \cite[8.331.3]{integrals} $\int_{0}^{x} u^{r-1} \diff u = (\frac{1}{r}) x^{r} = \frac{\Gamma(r)}{\Gamma(r+1)} x^{r}$ in 	\eqref{eq:appendix_zaf}  and apply the definition of Meijer's G-function to get \eqref{eq:cdf_prod_ris}.

Similarly, the  MGF of $Z_{i}$  $M_{Z_i}(s) = \mathbb{E}[e^{-sx}] = \int_{0}^{\infty} e^{-sx} f_{Z_{i}}(x) \diff x$ can be expressed as
\small
	\begin{align}\label{eq:mgf_prod_ris_appendix}
		&M_{Z_i}(s) = \frac{1}{2\pi \J} \sum_{l_{1},\cdots,l_{L}=1}^{P} \prod_{j=1}^{L} \psi_{j} v_{j}^{k_{j}} \zeta_{l_{j}} \bigg(C_{l_{j}}\bigg)^{-\phi_{j}} \int\limits_{\mathcal{L}}  \bigg(\prod_{j=1}^{L}C_{l_{j}}\bigg)^{r} \nonumber \\ &\bigg(\int_{0}^{\infty} e^{-sx} x^{r-1} \diff x\bigg) \Bigg[\frac{\prod_{j=1}^{L}\prod_{w=1}^{m} \Gamma\big(-r+\phi_{j}+b_{l_{j},w}\big) }{\prod_{j=1}^{L}\prod_{w=n+1}^{p} \Gamma\big(-r+\phi_{j}+a_{l_{j},w}\big) } \nonumber \\ &\frac{ \prod_{j=1}^{L}\prod_{w=1}^{n} \Gamma\big(1+r-\phi_{j}-a_{l_{j},w}\big)\prod_{j=1}^{L}\big(\Gamma\big(v_{j}-r\big)\big)^{k_{j}}}{ \prod_{j=1}^{L}\prod_{w=m+1}^{q} \Gamma\big(1+r-\phi_{j}-b_{l_{j},w}\big)\prod_{j=1}^{L}\big(\Gamma\big(1+v_{j}-r\big)\big)^{k_{j}}}\Bigg]\diff r  		
	\end{align}
\normalsize	
	 Substituting the inner integral solution using \cite[3.381.4]{integrals} as $\int_{0}^{\infty} e^{-sx} x^{r-1} \diff x = s^{-r} \Gamma(r)$ in  \eqref{eq:mgf_prod_ris_appendix}, we apply the definition of Meijer's G-function to get \eqref{eq:mgf_prod_ris}.

\section*{Appendix C: PDF and CDF of $Z$}
We apply the inverse Laplace transform of the MGF to find the PDF of $Z=\sum_{i=1}^N Z_i$ as
$f_{Z}(z) = \mathcal{L}^{-1} \prod_{i=1}^{N} M_{Z_i}(s)	$. Thus, we use \eqref{eq:mgf_prod_ris_appendix} and interchange the order of integration to get
\small
	\begin{align}\label{eq:pdf_ris_appendix}
	&f_{Z}(x) = \sum_{l_{1,1},\cdots,l_{1,L}=1}^{P}\cdots\sum_{l_{N,1},\cdots,l_{N,L}=1}^{P}\prod_{i=1}^{N} \prod_{j=1}^{L} \psi_{i,j} v_{i,j}^{k_{i,j}} \zeta_{l_{i,j}} \bigg(C_{l_{i,j}}\bigg)^{-\phi_{i,j}} \nonumber \\ &\Bigg(\bigg(\frac{1}{2\pi \J}\bigg)^{N} \int\limits_{\mathcal{L}_{i}} \bigg(\prod_{j=1}^{L} C_{l_{i,j}}\bigg)^{n_{i}} \Bigg[\frac{\prod_{j=1}^{L}\prod_{w=1}^{m} \Gamma\big(-n_{i}+\phi_{i,j}+b_{l_{i,j},w}\big) }{\prod_{j=1}^{L}\prod_{w=n+1}^{p} \Gamma\big(-n_{i}+\phi_{i,j}+a_{l_{i,j},w}\big) } \nonumber \\ &\frac{ \prod_{j=1}^{L}\prod_{w=1}^{n} \Gamma\big(1+n_{i}-\phi_{i,j}-a_{l_{i,j},w}\big)\prod_{j=1}^{L}\big(\Gamma\big(v_{i,j}-n_{i}\big)\big)^{k_{i,j}}}{ \prod_{j=1}^{L}\prod_{w=m+1}^{q} \Gamma\big(1+n_{i}-\phi_{i,j}-b_{l_{i,j},w}\big)\prod_{j=1}^{L}\big(\Gamma\big(1+v_{i,j}-n_{i}\big)\big)^{k_{i,j}}}\nonumber \\ &\Gamma(n_{i})\Bigg]\bigg(\frac{1}{2\pi \J} \int\limits_{\mathcal{L}}s^{-\sum_{i=1}^{N} n_i}e^{sx} \diff s\bigg)\diff n_i\Bigg)  			
\end{align}
\normalsize
where ${\cal{L}}_i$ is an infinite contour in the complex  $n_i$-plane such that the integrand in \eqref{eq:pdf_ris_appendix} has no singularities. The convergence conditions of multiple  contour integrals representing multi-variate Fox's H-function is presented in \cite{Hai1995TheCP}.
To solve the inner integral, we apply \cite[8.315.1]{integrals}:
\begin{align}\label{inner_eq}
	\int\limits_{\mathcal{L}} s^{\sum_{i=1}^{N} n_i} e^{sx} \diff s =& \bigg(\frac{1}{x} \bigg)^{1+\sum_{i=1}^{N} n_i} \frac{2\pi \J}{\Gamma(-\sum_{i=1}^{N}n_i)} 
\end{align}
Using \eqref{inner_eq} in \eqref{eq:pdf_ris_appendix} and applying  the definition of $N$-Multivariate Fox's H-function in \cite[A.1]{M-Foxh}, we get \eqref{eq:PDF_final}.

To derive the CDF, we use	$F_{Z}(z) = \mathcal{L}^{-1} \prod_{i=1}^{N} \frac{M_{Z_i}(s)}{s}$ in  \eqref{eq:mgf_prod_ris_appendix} to get 	
\small
\begin{align}\label{eq:cdf_ris_appendix}
	&F_{Z}(x) = \sum_{l_{1,1},\cdots,l_{1,L}=1}^{P}\cdots\sum_{l_{N,1},\cdots,l_{N,L}=1}^{P}\prod_{i=1}^{N} \prod_{j=1}^{L} \psi_{i,j} v_{i,j}^{k_{i,j}} \zeta_{l_{i,j}} \big(C_{l_{i,j}}\big)^{-\phi_{i,j}} \nonumber \\ &\Bigg(\bigg(\frac{1}{2\pi \J}\bigg)^{N} \int\limits_{\mathcal{L}_{i}} \big(\prod_{j=1}^{L} C_{l_{i,j}}\big)^{n_{i}} \Bigg[\frac{\prod_{j=1}^{L}\prod_{w=1}^{m} \Gamma\big(-n_{i}+\phi_{i,j}+b_{l_{i,j},w}\big) }{\prod_{j=1}^{L}\prod_{w=n+1}^{p} \Gamma\big(-n_{i}+\phi_{i,j}+a_{l_{i,j},w}\big) } \nonumber \\ &\frac{\prod_{j=1}^{L}\prod_{w=1}^{n} \Gamma\big(1+n_{i}-\phi_{i,j}-a_{l_{i,j},w}\big)\prod_{j=1}^{L}\big(\Gamma\big(v_{i,j}-n_{i}\big)\big)^{k_{i,j}}}{ \prod_{j=1}^{L}\prod_{w=m+1}^{q} \Gamma\big(1+n_{i}-\phi_{i,j}-b_{l_{i,j},w}\big)\prod_{j=1}^{L}\big(\Gamma\big(1+v_{i,j}-n_{i}\big)\big)^{k_{i,j}}} \nonumber \\ &\Gamma(n_{i})\Bigg]\bigg(\frac{1}{2\pi \J} \int\limits_{\mathcal{L}}s^{-1-\sum_{i=1}^{N} n_i}e^{sx} \diff s\bigg)\diff n_i\Bigg)  			
\end{align}
\normalsize

\begin{figure*}
	\begin{eqnarray}
		\label{eq:proof1_zaf}
		\int_{0}^{\infty} f_{Z}(z) dz  = \lim_{s\rightarrow 0} \prod_{i=1}^{N} \prod_{j=1}^{L} \frac{\rho_{i,j}^2v_{i,j}^{k_{i,j}}}{\Gamma(\alpha_{{\scriptscriptstyle F}}(i,j))\Gamma(\beta_{{\scriptscriptstyle F}}(i,j))}    H_{0,0:2L+\sum_{j=1}^{L}k_{1,j},2L+\sum_{j=1}^{L}k_{1,j}+1;\cdots;2L+\sum_{j=1}^{L}k_{N,j},2L+\sum_{j=1}^{L}k_{N,j}+1}^{0,0:L+1,2L+\sum_{j=1}^{L}k_{1,j};\cdots;L+1,2L+\sum_{j=1}^{L}k_{N,j}}\nonumber\\\left[\begin{array}{c}s\prod_{j=1}^{L} \frac{(\beta_{{\scriptscriptstyle F}}(i,j)-1)h_lA_0}{\alpha_{{\scriptscriptstyle F}}(i,j)}\\.\\.\\.\\s\prod_{j=1}^{L} \frac{(\beta_{{\scriptscriptstyle F}}(i,j)-1)h_lA_0}{\alpha_{{\scriptscriptstyle F}}(i,j)}\end{array} \middle\vert \begin{array}{c}
			-:\{\{(1-\alpha_{{\scriptscriptstyle F}}(i,j),1)\}_{j=1}^{L},\{(1-\rho_{i,j}^{2},1)\}_{j=1}^{L},\{\{(1-v_{i,j},1)\}_{1}^{k_{i,j}}\}_{j=1}^{L}\}_{i=1}^{N} \\		-:\{(0,1),\{(\beta_{{\scriptscriptstyle F}}(i,j),1)\}_{j=1}^{L},\{(-\rho_{i,j}^{2},1)\}_{j=1}^{L},\{\{(-v_{i,j},1)\}_{1}^{k_{i,j}}\}_{j=1}^{L},\}_{i=1}^{N}
		\end{array}\right] 	
	\end{eqnarray}
	\hrule 
\end{figure*}

We apply \cite[8.315.1]{integrals} to solve the inner integral in \eqref{eq:cdf_ris_appendix}:
\begin{align}\label{eq:inner_eq_2}
	\int\limits_{\mathcal{L}} s^{-1-\sum_{i=1}^{N} n_i} e^{sx} \diff s =& \bigg(\frac{1}{x} \bigg)^{-\sum_{i=1}^{N} n_i} \frac{2\pi \J}{\Gamma\big(1+\sum_{i=1}^{N}n_i\big)} 
\end{align}	
Using \eqref{eq:inner_eq_2} in \eqref{eq:cdf_ris_appendix}, we apply  the definition of $N$-Multivariate Fox's H-function in \cite[A.1]{M-Foxh} to get \eqref{eq:CDF_final} of Theorem 2.

We  validate the  derived PDF in \eqref{eq:PDF_final} by $\int_{0}^{\infty} f_{Z}(z) dz = 1$.  Using  the parameters of $\cal{F}$-distributed atmospheric turbulence from Table \ref{table:unified_param} in \eqref{eq:PDF_final}, we use the definition of Fox's H-function and interchange the order of integration to solve the inner integral using the final value theorem:
$	\int_{0}^{\infty} x^{-1-\sum_{i=1}^{N} x_i} dz = \lim_{s\rightarrow 0} (\frac{1}{s} )^{-\sum_{i=1}^{N} x_i} \Gamma (-\sum_{i=1}^{N} x_i)$. Thus, we get  \eqref{eq:proof1_zaf}.

Then, we use  \cite{Rahama2018} and apply  standard mathematical procedure for getting the limit of a function at $s\to 0$ in \eqref{eq:proof1_zaf}  to get a simplified expression, which results into
\begin{eqnarray}
	\int_{0}^{\infty} f_{Z}(z) dz =  \prod_{i=1}^{N} \prod_{j=1}^{L} \frac{\rho_{i,j}^2v_{i,j}^{k_{i,j}}}{\Gamma(\alpha_{{\scriptscriptstyle F}}(i,j))\Gamma(\beta_{{\scriptscriptstyle F}}(i,j))} \nonumber \\ \frac{\Gamma(\beta_{{\scriptscriptstyle F}}(i,j))\Gamma(\alpha_{{\scriptscriptstyle F}}(i,j))\Gamma(\rho_{i,j}^{2})\big(\Gamma(v_{i,j})\big)^{k_{i,j}}}{\Gamma(1+\rho_{i,j}^{2})\big(\Gamma(1+v_{i,j})\big)^{k_{i,j}}} =1		
\end{eqnarray}

\bibliographystyle{IEEEtran}
\bibliography{FSO,RIS_Letter,fso_fog}
\vspace{-10cm}
\begin{IEEEbiography}[{\includegraphics[width=1in,height=1.25in,clip,keepaspectratio]{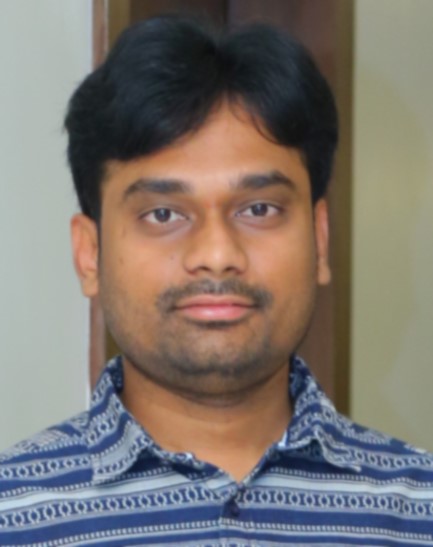}}]%
	{Vinay Kumar Chapala} received the B.Tech degree in Electronics and Communication Engineering from Jawaharlal Nehru Technology University, Hyderabad in 2010 and the M.Tech degree in Communications Engineering from Indian Institute of Technology, Delhi in 2013. He is  a Staff Engineer with Qualcomm India Pvt Ltd,  Bangalore.  He is currently pursuing  the PhD degree in communication  systems  with  the Department of Electrical and Electronics Engineering, Birla Institute of Technology and Science at Pilani, Pilani, India. His current research interests include signal processing and machine learning for wireless communications, and reconfigurable intelligent surfaces for wireless systems.
\end{IEEEbiography}
\vspace{-10cm}
\begin{IEEEbiography}[{\includegraphics[width=1.1in,height=1.35in,clip,keepaspectratio]{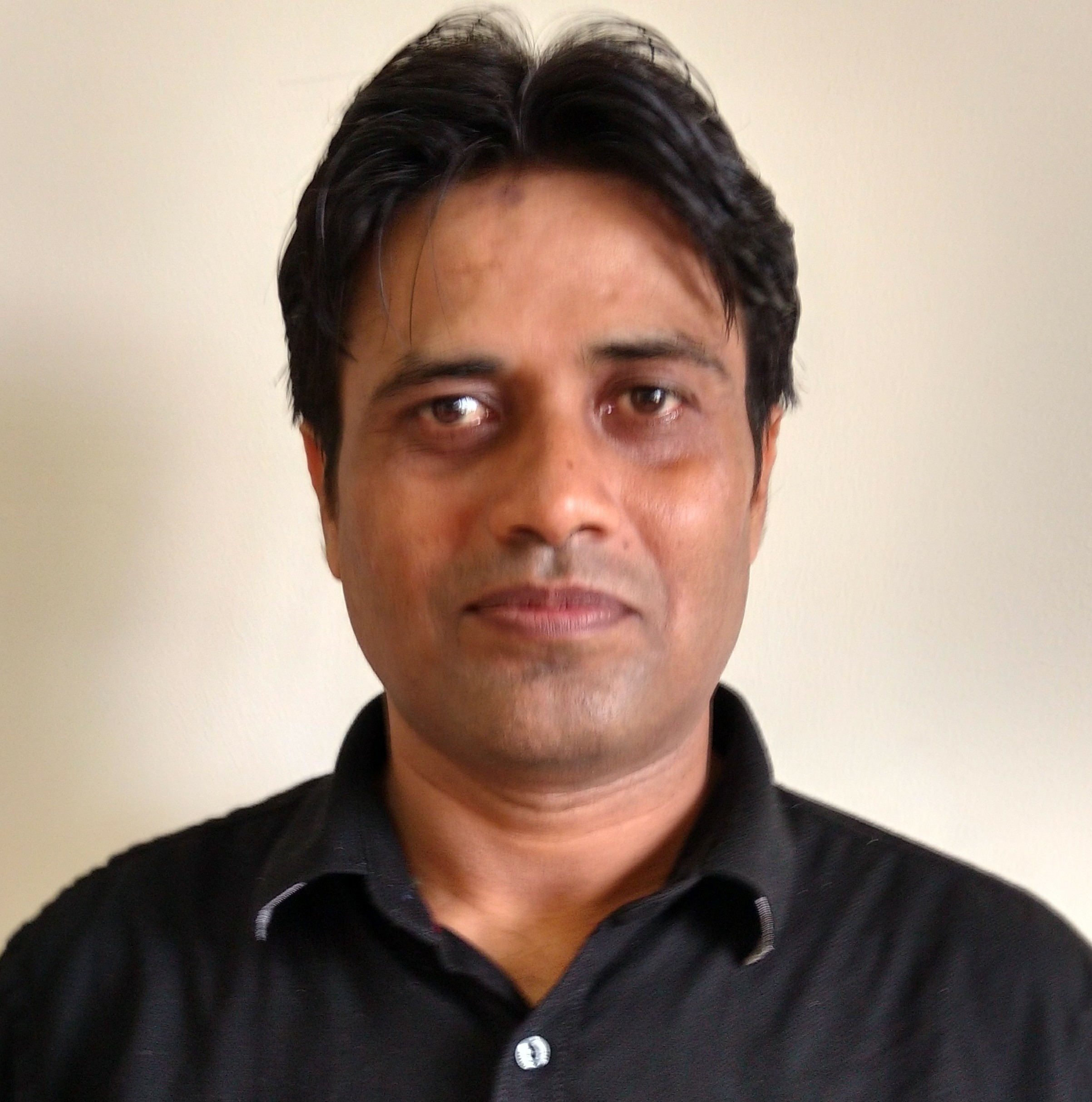}}]%
	{S.~M.~Zafaruddin } (Senior Member, IEEE) received the Ph.D.  degree in electrical engineering from IIT Delhi,  New Delhi, India, in 2013. From 2012 to 2015, he was with Ikanos Communications (now Qualcomm), Bangalore, India, working directly with  the CTO Office, Red Bank, NJ, USA, where he was involved in research and development for xDSL systems. From 2015 to 2018, he was a Post-Doctoral Researcher with the Faculty of Engineering, Bar-Ilan University, Ramat Gan, Israel, where he was involved in signal processing for wireline and wireless communications. He is currently an Assistant Professor with the Department of Electrical and Electronics Engineering, Birla Institute of Technology and Science at Pilani, Pilani, India. His current research interests include signal processing and machine learning for wireless and wireline communications, THz wireless technology, optical wireless communications, reconfigurable intelligent surface, distributed signal processing,  and resource allocation algorithms. He received the Planning and Budgeting Commission Fellowship for  Outstanding Post-Doctoral Researchers from China and India by the Council for Higher Education, Israel (2016–2018). He is also an Associate Editor of the IEEE ACCESS.
\end{IEEEbiography}

\end{document}